\documentclass[twocolumns]{IEEEtran}

 \usepackage{type1cm}        
\usepackage{multicol}        
\usepackage[bottom]{footmisc}

\usepackage{newtxtext}       %
\usepackage{newtxmath}   
\newcommand{\fg}{1}
\usepackage{cite}
\usepackage{verbatim}
\usepackage[pdftex]{graphicx}
\usepackage{amsfonts}
\usepackage{amsmath}
\usepackage{bbm}
\usepackage{float}
\usepackage{graphicx}
\usepackage{tikz}
\usepackage{xcolor}
\usepackage{multirow}
\usepackage{array}  
\usepackage{enumitem}
\usepackage{bbm}
\usepackage{color}
\usepackage{graphicx}
\usepackage{verbatim}
\usepackage{epstopdf}
\usepackage{cite}
\usepackage{tabulary}
\usepackage{mathtools}
\usepackage{enumitem}
\setlist{nolistsep}

\newcommand\indside[1]{\mathbbm{1}\left({#1}\right)}

\newcommand{\ie}{{\em i.e. } }

\newcommand{\tx}{\mathsf{t}}
\newcommand{\rx}{\mathrm{R}}
\newcommand{\rcvd}{\mathrm{r}}
\renewcommand{\L}{\mathrm{L}}
\newcommand{\N}{\mathrm{N}}

\def\home{\hbox{\kern3pt \vbox to13pt{}%
   \pdfliteral{q 0 0 m 0 5 l 5 10 l 10 5 l 10 0 l 7 0 l 7 5 l 3 5 l 3 0 l f
               1 j 1 J -2 5 m 5 12 l 12 5 l S Q }%
   \kern 13pt}}

\graphicspath{{figures/}}

\renewcommand{\ie}{\textit{i.e.,}~}
\newcommand{\subsubsubsection}[1]{\par\textit{#1}: }
\begin{document}

\title{Millimeter-wave and Terahertz Spectrum\\ for 6G Wireless}
\author{Shuchi Tripathi, Nithin V. Sabu, Abhishek K. Gupta, Harpreet S. Dhillon
%
\thanks{Shuchi Tripathi,   Nithin V. Sabu and  Abhishek K.  Gupta are with the Indian Institute of Technology Kanpur, India, Email: {\tt shuchi@iitk.ac.in,nithinvs@iitk.ac.in,gkrabhi@iitk.ac.in}.  Harpreet S. Dhillon is with Wireless@VT, Bradley Department of Electrical and Computer Engineering, Virginia Tech, Blacksburg, USA, Email: {\tt hdhillon@vt.edu}. 
This research is supported by the  Science and Engineering Research Board  (DST, India) under the grant SRG/2019/001459.

This is a preprint of the chapter that will appear in \cite{thebook}. Here are the complete details of the chapter: 

S. Tripathi, N. V. Sabu, A. K. Gupta, H. S. Dhillon, "Millimeter-wave and Terahertz Spectrum for 6G Wireless", in {\em 6G Mobile Wireless Networks}. Y. Wu, S. Singh, T. Taleb, A. Roy, H. S. Dhillon, M. R. Kanagarathinam, A. De, eds. Springer, 2021.
 }
}

\maketitle

\begin{abstract}

With the standardization of 5G, commercial millimeter wave (mmWave) communications has become a reality despite all the concerns about the unfavorable propagation characteristics of these frequencies. Even though the 5G systems are still being rolled out, it is argued that their gigabits per second rates may fall short in supporting many emerging applications, such as 3D gaming and extended reality. Such applications will require several hundreds of gigabits per second to several terabits per second data rates with low latency and high reliability, which are expected to be the design goals of the next generation 6G communications systems. Given the potential of terahertz (THz) communications systems to provide such data rates over short distances, they are widely regarded to be the next frontier for the wireless communications research. The primary goal of this chapter is to equip readers with sufficient background about the mmWave and THz bands so that they are able to both appreciate the necessity of using these bands for commercial communications in the current wireless landscape and to reason the key design considerations for the communications systems operating in these bands. Towards this goal, this chapter provides a unified treatment of these bands with particular emphasis on their propagation characteristics, channel models, design and implementation considerations, and potential applications to 6G wireless. A brief summary of the current standardization activities related to the use of these bands for commercial communications applications is also provided.
\end{abstract}
	

\section{Background and Motivation}
The standardization of 5G new radio (NR) was driven by the diverse throughput, reliability, and latency requirements of ever evolving ecosystem of applications that need to be supported by the modern cellular networks. Within 5G, these applications are categorized as enhanced mobile broadband (eMBB), ultra-reliable low latency communication (URLLC), and massive machine type communication (mMTC). Right from the onset, it was clear that a one-size-fits all solution may not work for all the applications because of which the recent generations of cellular systems have explored the use of advanced communications and networking techniques, such as network densification through the use of small cells, smarter scheduling, and multiple antenna systems for improved spectral efficiency, just to name a few. Perhaps the most striking difference of 5G from the previous generations of cellular systems is the acknowledgment that the {\em classical} sub-$6$ GHz spectrum is not going to be sufficient to support the requirements of the emerging applications. The millimeter wave (mmWave) spectrum naturally emerged as a potential solution. Although these bands were earlier thought to be unsuitable for the mobile operations due to their unfavorable propagation characteristics, the modern device and antenna technologies made it feasible to use them for commercial wireless applications \cite{TS2013a}. As a result, the 5G standards resulted in the birth of commercial mmWave communication.


Now, as we look into the future, it is evident that we are slowly moving towards applications, such as virtual and augmented reality, ultra-HD video conferencing, 3D gaming, and the use of wireless for brain machine interfaces, which will put even more strict constraints on the throughput, reliability, and latency requirements. With the advancement of device fabrication methods, it is also reasonable to expect that the nano-scale communications will see the light of the day soon. With the recent success of mmWave communication, it was quite natural for the researchers to start looking at the other unexplored bands of the radio frequency (RF) spectrum, primarily the terahertz (THz) band that lies above the mmWave band. The THz waves with enormous bandwidth can be used in many applications that require ultra-high data rates. This along with the existing sub-$6$ GHz and mmWave bands can help us achieve the true potential of many emerging applications. Further, owing to their small wavelength, they can also be used for micro and nano-scale communication. In the past, the use of THz bands was limited to imaging and sensing due to the unavailability of feasible and efficient devices that can work on these frequencies. However, with the recent advancements in THz devices, THz communication is expected to play a pivotal role in the upcoming generations of communication standards \cite{Sarieddeen2020}. 

The primary goal of this chapter is to equip readers with sufficient background about the mmWave and THz bands so that they are able to both appreciate the necessity of using these bands for commercial communications in the current wireless landscape and to reason the key design considerations for the communications systems operating in these bands. This is achieved through a systematic treatment of this topic starting with a detailed discussion of the propagation characteristics at these frequencies leading naturally to the discussion on channel models that capture these characteristics. Throughout this discussion, we carefully compare and contrast the propagation characteristics of these new bands with the better known sub-$6$ GHz cellular bands and explain how the key differences manifest in the channel models. Building on this background, we then explain the implications of these differences on the design considerations for mmWave and THz communications systems and their potential applications to 6G systems. The chapter is concluded with a brief discussion about the current standardization activities related to the use of these bands for commercial communications.


\section{Introduction to mmWave and  THz Spectrum}\label{sec:intro}


Until the 4G cellular standard, the commercial (cellular) communication was limited to the conventional bands up to $6$ GHz, which are now referred to as the sub-$6$ GHz cellular bands. However, there are many bands in the $6-300$ GHz range (with enormous bandwidths) that have been used for a variety of non-cellular applications, such as satellite communications, radio astronomy, remote sensing, radars, to name a few. Due to recent advancement in antenna technology, it has now become possible to use this spectrum for mobile communication as well. The frequency band from $30-300$ GHz with the wavelengths ranging from $1$ to $10$ mm is termed the {\em mmWave band} and offers hundreds of times more bandwidth compared to the sub-$6$ GHz bands. Although higher penetration and blockage losses are the major drawbacks of mmWave communication systems, researchers have shown that the same effects are helpful in mitigating interference in modern cellular systems, which exhibit dense deployment of small cells. This naturally results in a more aggressive frequency reuse and increased data security due to higher directionality requirement at the mmWave frequencies \cite{FK2012}. The mmWave frequencies from about $24$ GHz to about $100$ GHz are already being explored as a part of 5G standard. As we think ahead towards 6G and beyond systems, researchers have also started exploring the $0.1-10$ THz band, which is collectively referred to as the {\em THz band} (with the lower end of this spectrum being obviously of more interest for communications applications).


\subsection{Need for the mmWave and THz Bands}

It is well-known that the mobile data traffic has been exponentially increasing for more than a decade and this trend is expected to continue for the foreseeable future. With the penetration of wireless IoT devices in new verticals, such as supply chains, health care, transportation, and vehicular communications, this trend is further expected to accentuate. It is estimated that $9.5$ billion IoT devices are connected globally in $2019$ \cite{KLL2020}. The International Telecommunications Union (ITU) has further estimated that the number of connected IoT devices will rise to $38.6$ billion by $2025$ and $50$ billion by $2030$ \cite{IOT2015}, \cite{IOT2020}. Handling this data deluge and the massive number of IoT devices are two of the key design goals for 5G networks~\cite{dhillon2017wide}. Three possible solutions to meet these demands are to develop better signal processing techniques for improved spectral efficiency of the channel, the extreme densification of cellular networks, and the use of additional spectrum \cite{AG2020, JGA2014}. Various advanced techniques, such as carrier aggregation, coordinated multi-point processing, multi-antenna communications as well as novel modulation techniques have already been explored in the context of current cellular networks. The chances of getting orders of improvement from these techniques are slim. Likewise, network densification increases interference, which places fundamental limits on the performance gains that can be achieved with the addition of more base stations \cite{GupSabDhi2020,AndZDG2016}. The focus of this chapter is on the third solution, which is to use higher frequency bands. 

The amount of available spectrum at mmWave frequencies is very large when compared to sub-$6$ GHz frequencies ($\sim 50-100$ times). As the bandwidth appears in the pre-log factor of the achievable data-rate, mmWave communication can potentially achieve an order of magnitude higher data rate, which made it attractive for inclusion in the 5G standards. While 5G deployments are still in their infancy, emerging applications such as extended reality may require terabits-per-second (Tbps) links that may not be supported by the 5G systems (since the contiguous available bandwidth is less than $10$ GHz). This has created a lot of interest in exploring the THz band to complement the sub-$6$ GHz and mmWave bands in 6G and beyond systems \cite{Han2019,saad2019vision}.


\subsection{What can mmWave and THz Frequencies Enable?}

Larger bandwidths available in the mmWave spectrum make multi-gigabit wireless communication feasible, thus opening doors for many innovations \cite{FK2012}. For instance, the mmWave frequencies can enable wireless backhaul connections between outdoor base stations (BSs), which will reduce the land-acquisition, installation and maintenance costs of the fiber-optic cables, especially for ultra-dense networks (UDNs). Further, it enables to transform current ``wired'' data centers to completely wireless data centers with data-servers communicating over mmWave frequencies with the help of highly-directed pencil-beams. Another potential application is the in-boggy vehicle-to-vehicle (V2V) communication in high mobility scenarios including bullet trains and airplanes where mmWave communication systems together with sub-$6$ GHz systems have the potential of providing better data rates \cite{JGA2017}. 

Further, the THz spectrum consists of bands with available bandwidths of a few tens of GHz, which can support a data rates in the range of Tbps. The communication at THz is further aided by the integration of thousands of sub-millimeter antennas and lower interference due to higher transmission frequencies. It is therefore capable of supporting bandwidth-hungry and low latency applications, such as virtual-reality gaming and ultra-HD video conferencing. Other applications that will benefit from the maturity of THz communications include nano-machine communication, on-chip communications, internet of nano things (IoNT) \cite{Akyildiz2010}, and intra-body communication of nano-machines. It can also be combined with bio-compatible and energy-efficient bio-nano-machines communicating using chemical signals (molecules) \cite{Yang2020}. Such communication is termed \textit{molecular communication} \cite{Sabu2019}.


\begin{tiny}
\begin{table*}[ht!]
\caption{Available bands at the mmWave  \cite{AG2020, ITU2019} and THz  spectrum \cite{Kurner2020}.}
\begin{tabular}{| m{.15\textwidth} | p{.2\textwidth} | p{.57\textwidth} |}
\hline
Name & Specific bands & Remarks
\\
\hline
$26$ GHz band & $26.5-27.5$ GHz, $24.25 - 26.5$ GHz & Incumbent services: fixed link services, satellite Earth station services, and short-range devices. Earth exploration satellites and space research expeditions, inter-satellites, backhaul, TV broadcast distribution, fixed satellite Earth-to-space services and high altitude platform station (HAPS) applications.\\
\hline
$28$ GHz band & $27.5 - 29.5$ GHz,  $26.5 - 27.5$ GHz, & Proposed mobile communication. Incumbent services: Local multi-point distribution service (LMDS), Earth-to-space fixed-satellite service and Earth stations in motion (ESIM) application.   \\
\hline
$32$ GHz band & $31.0-31.3$ GHz, $31.8-33.4$ GHz & Highlighted as a promising band. Incumbent services: HAPS applications, Inter-satellite service (ISS) allocation. \\
\hline
$40$ GHz lower band & $37.0-39.5$ GHz, $39.5-40.5$ GHz & Incumbent services: Fixed and mobile satellite (space-to-Earth) and Earth exploration and space research satellite (space-to-Earth and Earth-to-space) services, HAPS applications. \\
\hline
$40$ GHz upper band & $40.5 - 43.5$ GHz & Incumbent services: Fixed and mobile satellite (space-to-Earth), broadcasting satellite services, mobile services, and radio astronomy. \\
\hline
$50$ GHz & $45.5 - 50.2$ GHz, $47.2 - 47.5$ GHz,  $47.9 - 48.2$ GHz, $50.4 - 52.6$ GHz & Incumbent services: Fixed non-geostationary satellite and international mobile telecommunication (IMT) services, HAPS applications. \\
\hline
{\color{black}$60$ GHz lower band} & $57.0 - 64.0$ GHz  &
Unlicensed operation for personal indoor services, device to device communication via access and backhaul links in the ultra-dense network scenario.\\
\hline
{\color{black}$60$ GHz upper band} & $64.0 - 71.0$ GHz & Upcoming generations of mobile standards with unlicensed status in UK and USA.  Incumbent services: The aeronautical and land mobile services.\\
\hline
{\color{black}$70/80/90$ GHz band} & $71.0 - 76.0$ GHz, $81.0 - 86.0$ GHz, $92.0 - 95.0$ GHz  & Fixed and broadcasting satellite services (space-to-Earth) services. Unlicensed operation for wireless device to device and backhaul communication services in the ultra-dense network scenario in the USA.\\
\hline
 $252 - 296$ GHz band& $252 - 275$ GHz,  $275 - 296$ GHz & Early proposal for land mobile and fixed service. Suitable for outdoor usage.\\
\hline
$306 - 450$ GHz band & $306 - 313$ GHz, $318 - 333$ GHz,  $356 - 450$ GHz &  Early proposal for land mobile and fixed service. Suitable for short range indoor communication.\\
\hline
\end{tabular}
\label{mmWaveTable}   
\end{table*}
\end{tiny}


\subsection{Available Spectrum} 

Due to the varying channel propagation characteristics and frequency-specific atmospheric attenuation, researchers have identified specific bands in mmWave/THz spectrum that are particularly conducive for the communications applications. In the world radiocommunication conference (WRC) 2015, ITU released a list of proposed frequency bands in between $24-86$ GHz range for global usage \cite{ITU2015}. The selection of these bands was done based on a variety of factors, such as channel propagation characteristics, incumbent services, global agreements, and the availability of contiguous bandwidth. WRC-2019 was focused on the conditions for the allocation of high-frequency mmWave bands dedicated to the 5G systems. A total of $17.25$ GHz of spectrum had been identified \cite{ITU2019}. For the implementation of future THz communication systems, WRC 2019 has also identified a total of $160$ GHz spectrum in the THz band ranging between $252$ to $450$ GHz. A brief description of these mmWave and THz bands are given in Table-\ref{mmWaveTable}. 

Although mmWave and THz bands have a huge potential for their usage in communication, there are significant challenges in their commercial deployments. In particular, communication in these bands suffer from poor propagation characteristics, higher penetration, blockage and scattering losses, shorter coverage range, and a need for strong directionality in transmission. These challenges have obstructed the inclusion of mmWave and THz bands in standards and commercial deployments until now. With the advancements in modern antenna and device technologies, it is now becoming feasible to use these bands for communications. However, there are still various design issues that need to be addressed before they can be deployed at a large scale \cite{JGA2017}, \cite{ED2020}. In this chapter, we will discuss the propagation characteristics of these bands in detail as well as the challenges involved in using them for communications applications. 


\section{Propagation at the mmWave and THz Frequencies}


\subsection{Differences from the Communication in Conventional Bands  }  

The communication at mmWave/THz frequencies differs significantly from the communication at conventional microwave frequencies. This is attributed to the following important factors. 


\subsubsection{Signal Blockage}


The mmWave/THz signals have a much higher susceptibility to blockages compared to the signals at the lower frequencies. The mmWave/THz communication relies heavily on the availability of line of sight (LOS) links due to very poor propagation characteristics of the non line of sight (NLOS) links \cite{FB2014}. For instance, these signals can be easily blocked by buildings, vehicles, humans, and even foliage. A single blockage can lead to a loss of $20-40$ dB. For example, the reflection loss due to glass for the mmWave signal is $3-18$ dB while that due to building material like bricks is around $40-80$ dB. Even the presence of a single tree amounts to a foliage loss of $17-25$ dB for the mmWave signals \cite{JGA2017, IH2018, SR2014, JGA2014}. Moreover, the mmWave/THz signals also suffer from the self-body blockage caused by the human users which can itself cause an attenuation of around 20-35 dB \cite{IH2018}. These blockages can drastically reduce the signal strength and may even result in a total outage. Therefore, it is of utmost importance to find effective solution to avoid blockages and quick handovers in case a link gets blocked. On the flip side, blockages, including self-body blockages, may also reduce interference, especially from the far off BSs \cite{Petrov2017}. Therefore, it is crucial to accurately capture the effect of blockages in the analytical and simulation models of mmWave/THz communications systems. 


\subsubsection{High Directivity} 

The second important feature of mmWave/THz communication is its high directivity. In order to overcome the severe path loss at these high frequencies, it is necessary to use a large number of antennas at the transmitter and/or receiver side \cite{FB2014}. Fortunately, it is possible to accommodate a large number of antennas in small form factors because antennas at these frequencies are smaller than those at traditional frequencies due to the smaller wavelengths. The use of a large antenna array results in a highly directional communication. High beamforming gain with small beamwidth increases the signal strength of the serving links while reducing the overall interference at the receivers. However, high directionality also introduces the {\em deafness problem} and thus higher latency. This latency occurs due to the longer beam search process which is a key step to facilitate the directional transmission and reception. This problem is aggravated in the high mobility scenario because both the user as well as the BSs suffer from excessive beam-training overhead. Therefore, new random access protocols and adaptive array processing algorithms are needed such that systems can adapt quickly in the event of blocking and handover due to high mobility at these frequencies \cite{MLA2020}. 
 
 
\subsubsection{Atmospheric Absorption} 

Electromagnetic (EM) waves suffer from transmission losses when they travel through the atmosphere due to their absorption by molecules of gaseous atmospheric constituents including oxygen and water. These losses are greater at certain frequencies, coinciding with the mechanical resonant frequencies of the gas molecules \cite{MM2005}. In mmWave and THz bands, the atmospheric loss is mainly due to water and oxygen molecules in the atmosphere, however, there is no prominent effect of atmospheric losses at the microwave frequencies. These attenuations further limit the distance mmWave/THz can travel and reduce their coverage regions. Therefore, it is expected that the systems operating at these frequencies will require much denser BS deployments. 


\subsection{Channel Measurement Efforts}  

Many measurement campaigns have been carried out to understand the physical characteristics of the mmWave frequency bands both in the indoor and outdoor settings. These measurement campaigns have focused on the study of path-loss, the spatial, angular, and temporal characteristics, the ray-propagation mechanisms, the material penetration losses and the effect of rain, snow and other attenuation losses associated with different mmWave frequencies. See Table \ref{MeasTable1} for a summary~\cite{TripGupMeasurementFile2020}.


\begin{tiny}
\begin{table*}[ht!]
\caption{MmWave channel measurements efforts for various environments.}
\begin{tabular}{
|p{3.5cm} | p{.76\textwidth} |}
\hline
 \textbf{Scenario/Environment} & \textbf{Measurement efforts}
\\
\hline

Indoor settings such as office room, office corridors, university laboratory. &
\begin{itemize}
\item Narrowband propagation characteristics of the signal, received power and bit error rate (BER) measurements \cite{ART1988}. 
\item Measurements of the fading characteristics/distribution \cite{allen199060}.
\item
Effects of frequency diversity on multi-path propagation \cite{allen1991frequency}. 
\item 
RMS delay spread measurement \cite{davies1991wireless}.
\item
Effects of transmitter and receiver heights on normalized received power for LOS and NLOS regions \cite{yang2005impact}.
\end{itemize}
\\
\hline
 Outdoor settings such as university campus, urban environments, streets, rural areas, natural environments, and grasslands. & 
 \begin{itemize}
 \item Comparison of the propagation mechanisms and fading statistics of the received signals \cite{allen1991outdoor}.  
 \item 
Channel impulse response, CDFs of received signal envelope and RMS delay spread \cite{daniele1994outdoor}. 
\item
Effects of multi-path scattering over foliage attenuation, mean and standard deviation of the path-loss \cite{wang2005enhanced}. 
\item
Impacts of rain attenuation on the link availability and signal depolarization \cite{fong2005measurement}.
\item
The path loss exponents and mean RMS delay spread of LOS and NLOS paths  \cite{ben2011millimeter}\cite{rappaport2012broadband}.
\item
Outdoor measurement over a distance of $5.8$ km at  $120$ GHz \cite{hirata2012}.
\item 
Effects of ground reflections and human shadowing on LOS path-loss measurements \cite{keusgen2014propagation}, \cite{weiler2014measuring}.
\end{itemize}
\\
\hline
High-speed train (HST) channel propagation measurements in the outdoor scenario. & \begin{itemize}
 \item The reflection and scattering parameters for the materials of the deterministic and random objects present in the HST environment. The verification of channel model in terms of path-loss, shadow-fading, power delay profile and small-scale fading \cite{guan2018towards1}.\end{itemize}\\
\hline
 Outdoor to indoor (O2I) propagation measurements. &  \begin{itemize}
 \item Effects of outdoor to indoor penetration losses on the number of multi-path components, RMS delay spread, angular spread and receiver beam-diversity  \cite{bas2019outdoor}. \end{itemize}\\
\hline
\end{tabular}
\label{MeasTable1}  
\end{table*}
\end{tiny}



%

Likewise, in \cite{Ma2018a}, measurements have been carried out to characterize THz wireless links for both indoor and outdoor environments. In the case of outdoor environments, \cite{Ma2018a} showed that interference from unintentional NLOS paths could limit the BER performance. The impact of weather on high capacity THz links was discussed in \cite{Federici2016}. The frequency ranges which are suitable for THz communication have been studied in \cite{Priebe2012, B.Heile}. In \cite{Guan2019}, intra-wagon channel characterization at $60$ GHz and $300$ GHz are done using measurements, simulations, and modeling.


\subsection{Propagation at mmWave and THz Frequencies}  \label{sec:propagation}

We now discuss key propagation characteristics of the mmWave and THz frequencies. 


\subsubsection{Atmospheric Attenuation} 
\begin{figure}[ht!]
	\centering
	\includegraphics[width=\fg\linewidth,trim=30 30 30 30,clip]{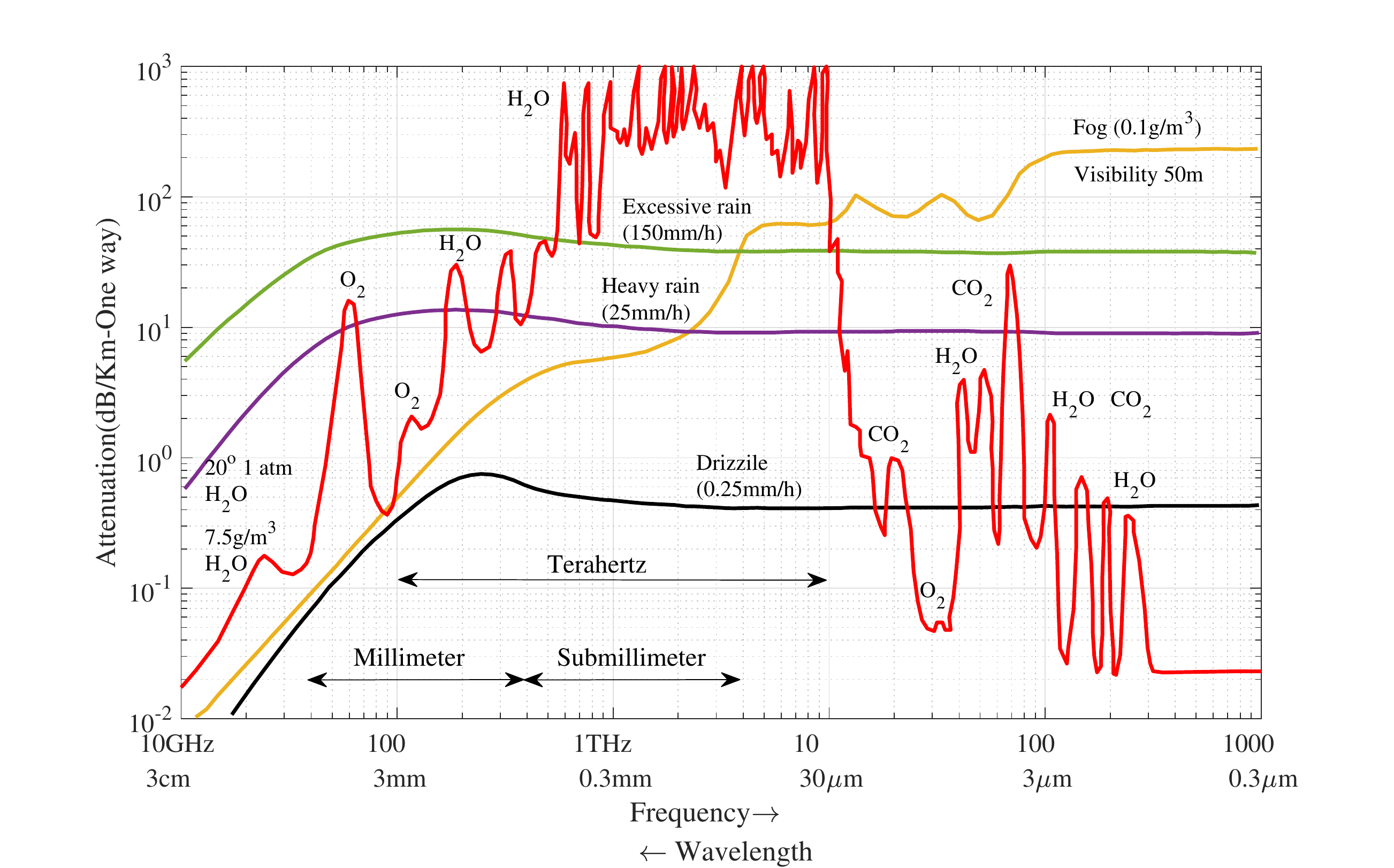}
	\caption{Variation of atmospheric absorption  due to various factors  with respect to frequencies in the band 10GHz-1000THz. This figure is reproduced using data from \cite{Lettington2002}.}
	\label{fig:abs2}
\end{figure}

The atmospheric attenuation is caused by the vibrating nature of gaseous molecules when exposed to the radio signals. Molecules with sizes comparable to the wavelength of EM waves excite when they interact with the waves, and these excited molecules vibrate internally. As a result of this vibration, a part of the propagating wave's energy is converted to kinetic energy. This conversion causes loss in the signal strength \cite{Jornet2011}. The rate of absorption depends upon the temperature, pressure, altitude and the operating carrier frequency of the signal. At lower frequencies (sub-$6$ GHz), this attenuation is not significant. But, higher frequency waves undergo significant attenuation since their wavelength becomes comparable to the size of dust particles, wind, snow, and gaseous constituents. The two major absorbing gases at mmWave frequencies are oxygen ($\mathrm{O_2}$) and water vapor ($\mathrm{H_2O}$). As seen in Fig. \ref{fig:abs2}, the peaks of $\mathrm{O_2}$ absorption losses are observed at $60$ GHz and $119$ GHz which are associated with a loss of $15$ dB/km and $1.4$ dB/km, respectively. Similarly, the peaks of $\mathrm{H_2O}$ absorption losses are observed at $23$ GHz, $183$ GHz and $323$ GHz, which are associated with a loss of $0.18$ dB/km, $28.35$ dB/km and $38.6$ dB/km, respectively. Similarly, $380$ GHz, $450$ GHz, $550$ GHz, and $760$ GHz frequency bands also suffer a higher level of attenuation. However, for short-distance transmission, the combined effects of these atmospheric losses on mmWave signals is not significant \cite{IH2018}. THz communication is even more prone to the atmospheric effects in the outdoor environment. We can see that the spectrum between $600$ and $800$ GHz suffers $100$ to $200$ dB/km attenuation which is $\approx$10-20 dB over the distance of approximately $100$ m \cite{Rappaport2019}. The absorption process can be described with the help of Beer-Lambert's law which states that the amount of radiation of frequency $f$ that is able to propagate from a transmitter to the receiver through the absorbing medium (termed the transmittance of the environment) is defined as \cite{Kokkoniemi2015}
\begin{equation}
	\tau (r,f)=\frac{P_{rx}(r,f)}{P_{tx}(f)}=\exp(-\kappa_a(f) r),\label{eq:molabs}
\end{equation}
where $ P_{rx}(r,f) $ and $ P_{tx}(f) $ are the received and transmitter power, and $r$ is the distance between the transmitter and the receiver. Here, $ \kappa_a(f) $ denotes the absorption coefficient of the medium. The $ \kappa_a(f) $ is the sum of the individual absorption coefficient of each gas constituent, which depends on its density and type \cite{Jornet2011}.

\subsubsection{Rainfall Attenuation} 


The wavelengths of the mmWave spectrum range between $1$ to $10$ millimeters, whereas the average size of a typical raindrop is also in the order of a few millimeters. As a result, the mmWave signals are more vulnerable to blockage by raindrops than conventional microwave signals. The light rain (say, $2$ mm/hr) imposes a maximum loss of $2.55$ dB/km whereas the heavy rain (say, $50$ mm/hr) imposes a maximum loss of $20$ dB/km. In tropical regions, a monsoon downpour at $150$ mm/hr has a maximum attenuation of $42$ dB/km at frequencies over $60$ GHz. However, in the lower bands of the mmWave spectrum, such as the $28$ GHz and $38$ GHz bands, lower attenuations of around $7$ dB/km are observed during heavy rainfall, which drops to $1.4$ dB for the coverage range of up to $200$ m. Thus by considering short-range communications and lower bands of mmWave spectrum, the effect of rainfall attenuation can be minimized \cite{IH2018}.


\subsubsection{Blockage}

\begin{enumerate}
\item[(i)] \textit{Foliage attenuation:} The presence of vegetation can cause further attenuation at mmWave/THz frequencies. The severity of foliage attenuation depends on the carrier frequency and the depth of vegetation. For example, the foliage attenuation loss of $17$ dB, $22$ dB and $25$ dB are observed at $28$ GHz, $60$ GHz and $90$ GHz carrier frequencies, respectively \cite{IH2018}.

\item[(ii)] \textit{Material penetration losses:} The mmWave and higher frequencies cannot propagate well through obstacles like room furniture, doors and walls. For example, a high penetration losses of $24.4$ dB and $45.1$ dB was observed at $28$ GHz signal when penetrating through two walls and four doors, respectively \cite{IH2018}. The higher penetration losses limit the coverage region of the mmWave transmitter in the indoor-to-outdoor and outdoor-to-indoor scenarios.
\end{enumerate}



A LOS probability model can be used to incorporate the effects of static blockages  on the channel. This model assumes that a link of distance $d$ will be LOS with probability $p_{\L}(d)$ and NLOS otherwise. The expressions of $p_{\L}(d)$ are usually obtained empirically for different settings. For example, for the urban macro-cell (UMa) scenario \cite{haneda20165g}
\begin{align*}
	p_{\L}(d) = \min\left(\frac{d_1}{d}, 1\right)\left(1 - e^{-\frac{d}{d_2}}\right) + e^{-\frac{d}{d_2}},
\end{align*}
where $d$ is the 2D distance in meters and $d_1$ and $d_2$ were the fitting parameters equal to $18$ m and $63$ m, respectively. 
The same model is also applicable for the urban micro-cell (UMi) scenario, with $d_2=36$ m. There are some variations in the LOS probability expressions across different channel measurement campaigns and environments. For example, the LOS probability model developed by NYU \cite{samimi2015probabilistic} is 
\begin{align*}
	p_{\L}(d) = \left(\min\left(\frac{d_1}{d}, 1\right)\left(1 - e^{-\frac{d}{d_2}}\right) + e^{-\frac{d}{d_2}}\right)^2.
\end{align*}
where the fitting parameters $d_1$ and $d_2$ were equal to $20$ m and $160$ m, respectively. 

These empirical models can be justified theoretically. In \cite{bai2014analysis}, a cellular network with random rectangular blockages was considered where blockages were modelled using the Boolean process and it was shown that LOS probability is given as
\begin{align*}
	p_{\L}(d) &=  e^{-\beta d}, & \text{where }
	\beta &= \frac{2\mu(\mathbb{E}[W] + \mathbb{E}[L])}{\pi},
\end{align*}
where $L$ and $W$ are the length and width of  a typical rectangular blockage and $\mu$ is the density of blockages. A different blockage model known as the {\em LOS ball model} was introduced in \cite{TB2014b} which assumes that all links inside a fixed ball of radius $R_B$ are LOS, \ie
\begin{align*}
	p_{\L}(d) &= \mathbb{I}(d < R_B), & \text{where } R_B = \frac{\sqrt{2}\mu\mathbb{E}[L]}{\pi},
\end{align*} 
which can also be used in the analysis of mmWave cellular networks.



\subsubsection{Human Shadowing and Self Blockage} 

As discussed earlier, propagation at mmWave/THz frequencies can suffer significant attenuation due to the presence of humans including the self-blockage from the user equipment itself. In \cite{gapeyenko2016analysis}, human body blockages were modeled using a Boolean model in which humans are modeled as 3D cylinders with centers forming a 2D Poisson point process (PPP). Their heights were assumed to be normally distributed. In indoor environments, human blockages have also been modeled as 2D circles of fixed radius $r$ with centers forming a PPP ($\mu$) \cite{KV2016}. The LOS probability for a link of length $d$ in this case comes out to be
\begin{align*}
	p_{\L} = 1 - e^{-\mu(rd+\pi r^2)}
\end{align*}
 
The self-blockage of a user can also be modeled using a 2D cone of angle $\delta$ (which is determined by the user equipment width and user to equipment distance), such that all BSs falling in this cone are assumed to be blocked \cite{bai2014}. 

%
%


\subsubsection{Reflections and Scattering} 
Consider an EM wave impinging on a surface. If the surface is smooth and electrically larger than the wavelength of the wave, we see a single reflection in a certain direction.
 The fraction of the incident field that is reflected in the specular direction is denoted by the reflection coefficient of the smooth surface, termed $ \Gamma_s$, which also accounts for the penetration loss. The reflected power is thus
\begin{align*}
	\mathrm{\overline{P}_R}=\mathrm{P}\Gamma_s^2,
\end{align*}
where $P$ is the power of incident wave. However, if the surface is rough, the wave gets scattered into many directions in addition to a reflected component in the specular direction. This phenomenon is known as \textit{diffuse scattering} \cite{Ju2019}, which is also exhibited by the mmWave/THz signals. As discussed next in detail, this behavior is attributed to the smaller wavelengths, which are comparable to the size of small structural features of the buildings surfaces. 

%
%
%

%
%

Most importantly, whether a surface will be perceived smooth or rough depends upon the incident wave's properties. 
%
The {\em Rayleigh criteria} can be used to determine the smoothness or roughness of a surface based on the critical height associated to the wave $h_c$, which is given as 
\cite{Ju2019}
\begin{align*}
	h_c=\frac{\lambda}{8\cos\theta_i},
\end{align*}
where $ h_c $ depends on the incident angle $ \theta_i $ and wavelength $ \lambda $. Let the minimum-to-maximum surface protuberance of the given surface be denoted by $h_0$, while the RMS height of the surface is 
$h_{\text {rms }}$. Then, if $ h_0<h_c $, the surface can be considered smooth, and if $ h_0>h_c $, the surface can be considered rough for the particular wave with wavelength $\lambda$. 
This implies that as $\lambda$ decreases, the same surface which was smooth at higher $\lambda$, may start becoming rough. Therefore, at lower frequencies, reflection phenomenon is significant, while scattering is negligible as most surfaces are smooth compared to the wave. As a result, reflections are more prominent in the lower mmWave bands while the scattering is moderate. However, as we go higher in frequency to the THz bands, scattering becomes significant since the roughness in the surface of building walls and terrains becomes comparable to the carrier wavelength. As a result, the scattered signal components at THz are more significant compared to the reflected paths.

\begin{figure}[t]
	\centering
	\includegraphics[width=\fg\linewidth,clip,trim=0 100 0 100]{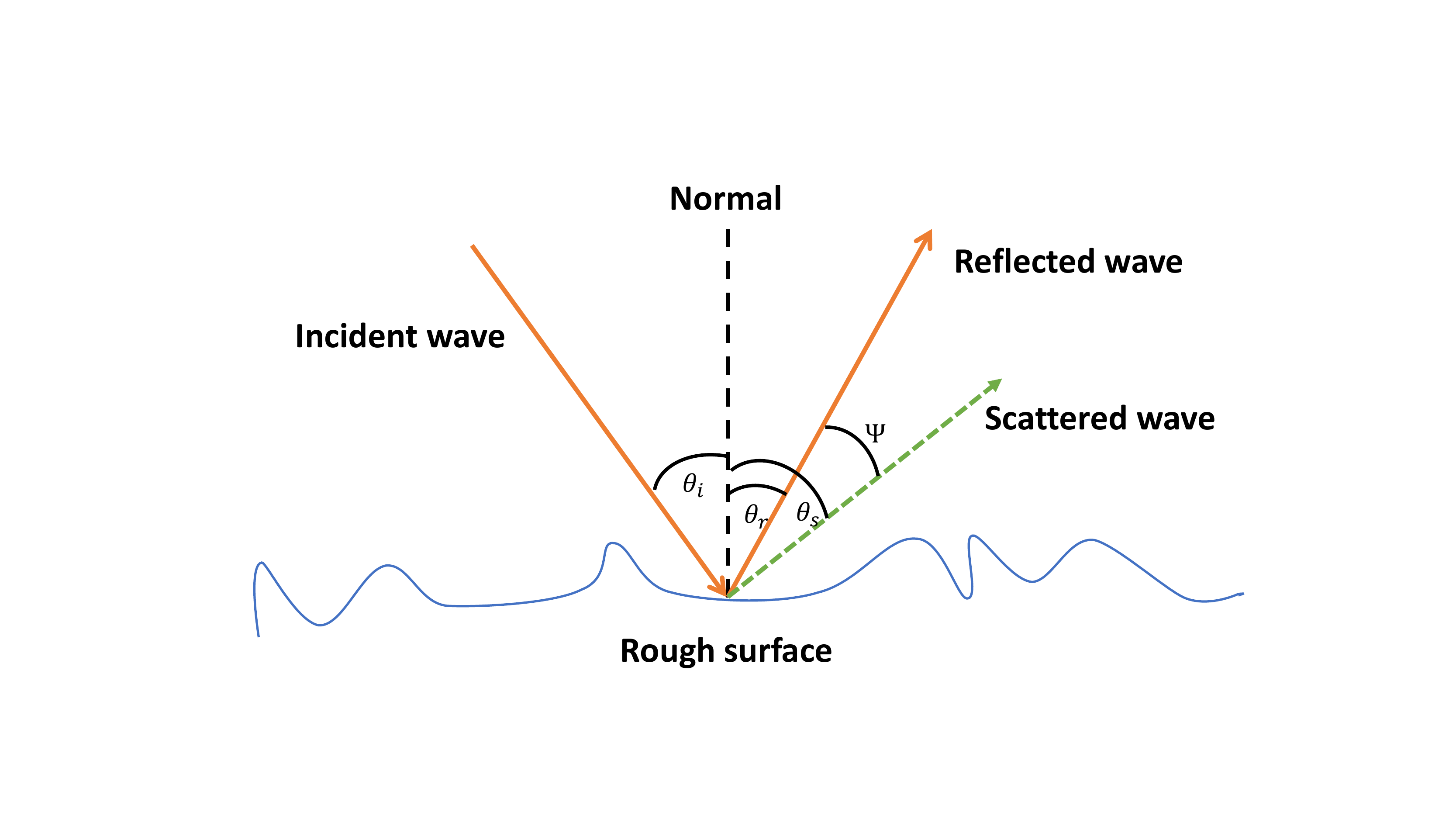}
	\caption{Schematic diagram of a radio wave incident at a surface. $\theta_i$ is the incident angle, $\theta_r$ is the reflected angle, $ \theta_s $ is the scattered angle and $ \Psi $ is the angle between reflected and scattered waves.}
	\label{fig:scattering}
\end{figure}

For rough surfaces, scattering results in additional loss in the reflected wave, if there is one. Therefore, the scattering loss factor (denoted by $\rho $) has to be considered to obtain the reflection coefficient $ \Gamma $ of a rough surface 
\cite{Ju2019}
\begin{align*}
	\Gamma=\rho\Gamma_s, &&\text{with }
	\rho\approx \exp \left[-8\left(\frac{\pi h_{\mathrm{rms}} \cos \theta_{i}}{\lambda}\right)^{2}\right].
\end{align*}
Therefore, the scattered power from this surface is given by
\begin{align}
	\mathrm{\overline{P}_S}=\mathrm{P}\left(1-\rho^2\right)\Gamma_s^2,\label{eq:tot_scat_power}
\end{align}
and the reflected power is given by
\begin{align}
	\mathrm{\overline{P}_R}=\mathrm{P}\Gamma^2=\mathrm{P}\rho^2\Gamma_s^2. \label{eq:tot_refl_power}
\end{align}
The  fraction of the incident wave that is scattered is represented by scattering coefficient $ S^2 $. The scattering coefficient $ S^2 $ is given by
\begin{align*}
	S^2=\frac{\mathrm{\overline{P}_S}}{\mathrm{P}}=\left(1-\rho^2\right)\Gamma_s^2.
\end{align*}
There are various models to characterize the variation of scattering power with scattering direction. One of the widely used models is the directive scattering (DS) model, 
which states that the main scattering lobe is steered in the general direction of the specular reflected wave ($ \theta_r $ in Fig. \ref{fig:scattering}) and the scattered power in a direction $\theta_s$ is 
\begin{align*}
	\mathrm{P_S}(\theta_s)&\propto\left(\frac{1+\cos(\theta_s-\theta_r)}{2}\right)^{\alpha_R},
\end{align*}
where $\alpha_R $ represents the width of the scattering lobe. 
In \cite{Jarvelainen2012}, DS model is used to model the propagation of a 60GHz wave in the hospital room. The DS model  was found to agree with rural and suburban buildings scattering when validated with $1.29$ GHz propagation measurements \cite{Degli-Esposti2007}. 

We can now compute the scattered power at a receiver from a transmitter located at $r_i$ distance away from the surface. From the Friis equation and \eqref{eq:tot_scat_power}, the total scattered power from the surface can be expressed as
\begin{align*}
	\mathrm{\overline{P}_S}=S^2A_s\frac{\mathrm{P}_\tx G_t}{4\pi r_i^2},
\end{align*}
where $\mathrm{P}_\tx$ is the transmitted power, $ G_t $ is the transmitter antenna gain, and $A_s$ is the effective aperture of the scattering surface. Now from DS model, the scattered power $\mathrm{{P}_S}$ at a distance $ r_s $ in the direction $\theta_s$,  is
\begin{align*}
	\mathrm{{P}_S}(\theta_s)&=\mathrm{P_{S0}}\left(\frac{1+\cos(\theta_s-\theta_r)}{2}\right)^{\alpha_R}
\end{align*}
where 
$ \mathrm{P_{S0}}$ is the maximum scattered power given as
\begin{align*}
	\mathrm{P_{S0}}&=\frac{\mathrm{\overline{P}_S}}{r_s^2\int \int \left(\frac{1+\cos(\theta_s-\theta_r)}{2}\right)^{\alpha_R}\mathrm{d}\theta_s\mathrm{d}\phi_s}
\end{align*}
If we define $F_\alpha=\int \int \left(\frac{1+\cos(\theta_s-\theta_r)}{2}\right)^{\alpha_R}\mathrm{d}\theta_s\mathrm{d}\phi_s$, then 
\begin{align*}
	\mathrm{P_{S0}}
	=\frac{\mathrm{\bar{P}_S}}{r_s^2F_\alpha}
	&=S^2A_s\frac{\mathrm{P}_\tx G_t}{4\pi r_i^2}\frac{1}{r_s^2F_\alpha}.
\end{align*} 
Hence, the received power at the receiver located at an angle $\theta_r$ and a distance $r_s$ from the surface, is given as
%
\begin{align}
	\mathrm{p}_\rcvd&=\mathrm{P_S}(\theta_s)\times \text{Effective antenna aperture}	=\mathrm{P_S}\frac{\lambda^2}{4\pi}G_r\nonumber\\
	&=S^2A_s\frac{\mathrm{P}_\tx G_t}{4\pi }\frac{1}{r_i^2r_s^2}\frac{\lambda^2}{4\pi}G_r\frac{1}{F_\alpha}\left(\frac{1+\cos(\theta_s-\theta_r)}{2}\right)^{\alpha_R},
\end{align}
where 
 $ G_r $ is the receiver antenna gain. 
The model can also be extended to consider the backscattered lobe.


\subsubsection{Diffraction} 
Owing to its short wavelength, in mmWave/THz frequencies, diffraction will not be as prominent as it is at microwave frequencies \cite{Rappaport2019}. In these frequencies, NLOS has significantly less power compared to that of the LOS path
\cite{KulCorrectionFactor2018}. However, it may be possible to establish THz links in the shadow of objects with the help of diffraction \cite{Kokkoniemi2016}. 


\subsubsection{Doppler Spread} 
Since the Doppler spread is directly proportional to the frequency and the speed of users, it  is significantly higher at mmWave frequencies than the sub-6 GHz frequencies. For example, the Doppler spread at $30$ GHz and $60$ GHz is $10$ and $20$ times higher than at $3$ GHz \cite{IH2018}. 


\subsubsection{Absorption Noise} Along with attenuation in the signal power, molecular absorption causes the internal vibration in the molecules which results in the emission of EM radiation at the same frequency as that of the incident waves that provoked this vibration. Due to this, molecular absorption introduces an additional noise known as absorption noise. Since absorption is significant in the THz bands, absorption noise is included in the total noise as an additional term. It is generally modeled using an equivalent noise temperature of the surroundings caused by the molecular absorption \cite{Kokkoniemi2015}.


\subsubsection{Scintillation Effects} \textit{Scintillation} refers to the rapid fluctuation in the wave's phase and amplitude due to the fast local variation in the refractive index of the medium through which the wave is travelling. Local variation in temperature, pressure, or humidity causes small refractive index variations across the wavefront of the beam which can destroy the phase front, and the beam cross-section appears as a speckle pattern with a substantial local and temporal intensity variation in the receiver. Infrared (IR) wireless transmission distance is limited by scintillation effects \cite{Federici2010}. The result of scintillation on practical THz communication is smaller than the IR beams. The THz waves traveling close to the surface of the earth may be influenced by the atmospheric turbulence \cite{Bao2012}. However, the extent to which scintillation effects impact the THz bands is still not well understood.



\subsection{Beamforming and Antenna Patterns}


In multiple inputs and multiple outputs (MIMO) systems, beamforming is used to focus a wireless signal towards a specific receiver (or away from certain directions to avoid interfering with devices in those directions). 
The gain thus achieved in the signal to noise ratio (SNR) at the intended receiver is called the {\em beamforming gain}, which is essential in mmWave systems to ensure reliable reception. Traditional MIMO systems were based on the \textit{digital beamforming}, where each element in the antenna array has its separate digital-to-analog (D/A) conversion unit and the RF chain. However, fully digital beamforming is not suitable for mmWave frequencies due to many-fold increase in the number of antenna elements which not only increases the cost of the overall system but also the substantial power consumption \cite{JGA2017}. Further the power consumption generally scales linearly with the sampling rate and exponentially with the number of bits per samples \cite{JGA2017, IH2018, SR2014,JGA2014}.


In order to lower the power consumption, \textit{analog beamforming} has been proposed for mmWave systems where  a single RF chain is shared by all antenna elements. Each antenna is fed with the phase shifted version of the same transmit signal where phase shift is determined according to the beamforming direction. However, such tranmission is limited to a single stream and single user transmission/reception. To enable multi-user/multi-stream transmission for mmWave networks \cite{JGA2017, IH2018, SR2014,JGA2014}, {\em hybrid beamforming} has been proposed in which more than one RF chains are used. The hybrid beamforming architectures are broadly classified into two types, the \textit{fully connected hybrid beamforming architecture}, where each RF chain is connected to all antennas and the \textit{partially connected hybrid beamforming architecture}, where each RF chain is connected to a subset of antenna elements. Clearly, hybrid beamforming provides a tradeoff between low-complexity but restrictive analog beamforming and the high-complexity but most flexible fully digital beamforming.

\subsubsection{Analog Beamforming Patterns}
Due to analog beamforming, the effective gain in the received signal can be computed using the transmitter and receiver antenna patterns which represents the gain in different directions around the antenna array (e.g., see \eqref{eq:mmwavechannel}).  
Various antenna patterns have been proposed in the literature to aid the evaluation of mmWave systems. Some examples are discussed below.

\subsubsubsection{Uniform linear array (ULA) model}
For the antenna element spacing $d$ and signal wavelength $\lambda$, the antenna gain of an $N$-array ULA  \cite{CAB2016} is
\begin{align}
	G_{\text{act}} (\phi) = \frac{\sin^2(\pi N \phi)}{N^2 \sin^2 (\pi \phi)}, \label{ulaeq}
\end{align}
where  $\phi = \frac{d}{\lambda} \cos \theta$ is the cosine direction corresponding to the spatial angle of departure (AoD), $\theta$, of the transmit signal.
In order to avoid the grating lobes at mmWave frequencies, the antenna element spacing $d$ is generally kept to be half of the wavelength. Since the spatial angle $\phi$ depends on $d$, we can use the approximation $\sin (\pi \phi) \simeq \pi \phi$ in the denominator. Therefore the array gain function in (\ref{ulaeq}) can be approximated as a squared sinc-function
\begin{align}
	G_{\text{sinc}} (\phi) \triangleq \frac{\sin^2 (\pi N \phi)}{(\pi N \phi )^2}.
\end{align}
This \textit{sinc antenna pattern} has been widely used for the numerical analysis in antenna theory. Authors in \cite{XY2017} have verified the accuracy of tight lower bound provided by sinc antenna model for the actual antenna pattern that makes it highly suitable for the network performance analysis of the mmWave systems. 

\begin{figure} [t]
	\centering
	\includegraphics[width=0.66\fg\linewidth]{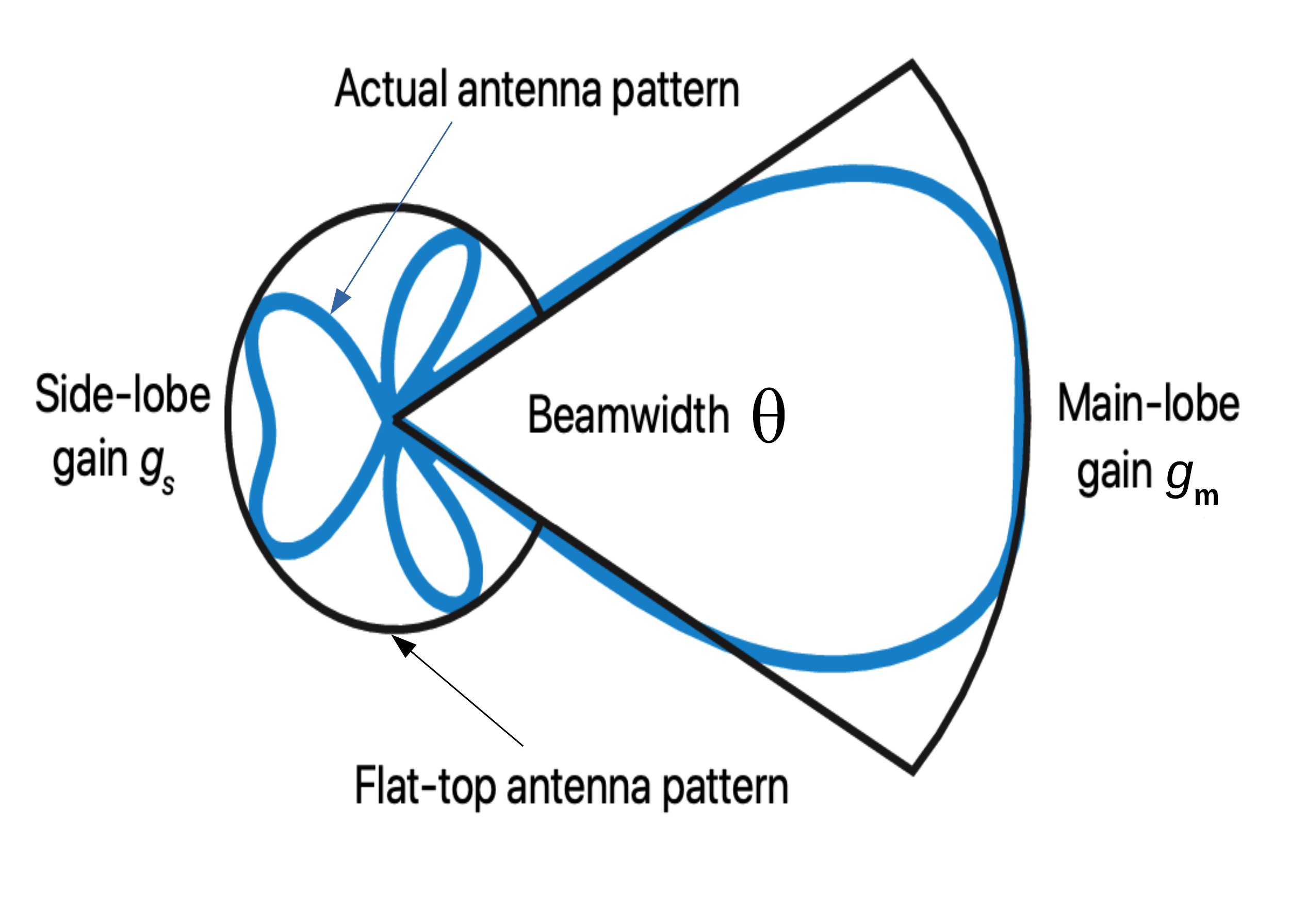}
	\caption{The sectorized antenna model \cite{JGA2017} which provides analytical tractability in the system level evaluations of the mmWave systems.} \label{flat}
\end{figure}

\subsubsubsection{Sectorized antenna model}
To maintain the analytical tractability in the network coverage analysis, many researchers approximate the actual antenna pattern with the \textit{flat-top antenna pattern}, also known as the \textit{sectorized antenna model} (see Fig. \ref{flat}).  
In this model, the array gains within the half-power beam-width (HPBW) $\theta_{\text{3dB}}$ are approximated to the maximum main-lobe gain $g_m$ while the array gains corresponding to the remaining AoDs are approximated to the first side-lobe gain $g_s$ of the actual antenna pattern \cite{TB2014b}. Hence, the gain in the direction $\theta$ is given as
\begin{align}
	G_{\text{Flat}}(\theta) = \begin{cases}
			g_m \qquad & \text{if} \ \theta \in [- \theta_{\text{3dB}},\ \theta_{\text{3dB}}] \nonumber \\
			g_s  \qquad & \text{otherwise}.
				\end{cases}
\end{align}
Thus, the flat-top antenna pattern models the continuously varying actual antenna array gains using the fixed main-lobe and side-lobe gains. For highly dense network scenarios, the aggregated interference from the side lobes is significant because of which the term $g_s$ must not be ignored in the analysis. This model is limited in its ability to fit arbitrary antenna patterns and is not suitable to analyze beam mis-alignments. 

\subsubsubsection{Multi-lobe antenna model}
In order to generalize the flat-top antenna pattern, a \textit{multi-lobe antenna model} was proposed in \cite{WL2015} where there are  $K$ number of lobes, each with a constant gain. The array gain and the width of each lobe are obtained by minimizing the error function between the multi-lobe pattern and the actual antenna pattern. 
%
%
The limitation of the model includes the lack of the roll-off characteristic of the actual antenna pattern because of which the predicted analytical performance of the network may deviate from the actual network performance \cite{MDR2016}.

\subsubsubsection{Gaussian antenna model}
The Gaussian antenna model is proposed in order  to capture the effects of roll-off in actual antenna pattern which generally occur due to small perturbations and misalignment between the receiver and transmitter \cite{AM2014,AT2015}. The antenna gain for this model is given as
\begin{align}
	G_{\text{Gaussian}}(\theta) = (g_m - g_s) e^{-\eta \theta^2} + g_s.
\end{align}
where $g_m$ is the maximum main-lode gain which occurs as $\theta = 0$, $g_s$ is the side-lobe gain and $\eta$ is a parameter that controls the $3$ dB beam-width. 

\subsubsubsection{Cosine antenna model}
The antenna pattern for \textit{cosine antenna pattern} is given as \cite{XY2017} 
\begin{align*}
	G_{\cos} (\theta) = 
			\cos^2 \left( \frac{\pi N}{2} \theta \right) 
			\indside{ \lvert \theta \rvert \leq \frac{1}{N}}.
\end{align*}
This model can be extended to include multiple lobes \cite{DN2018} to give additional flexibility. 
%


\subsubsection{Antenna Patterns for Multi-user/stream Transmission}
The above discussion can be extended to include the hybrid beamforming supporting multi-stream or multi-user transmission. As a {\em layered} technique, hybrid beamforming can be seen as linear combination of the digital and analog beamforming. Hence the effective antenna pattern in case of multi user or multi- stream transmission will consists of individual analog beam patterns, one for each stream or user. 

\subsubsection{THz Beamforming} 

The limited transmission range of THz waves can be extended somewhat via very dense ultra-massive multiple-input multiple-output (UM-MIMO) antenna systems. Since the number of antennas that can fit into the same footprint increases with the square of the wavelength, the THz systems can accommodate even larger number of antenna elements than mmWave systems. This large array of compact antennas results in highly focused beams (pencil beams) of high gain that aids in increasing the transmission distance.

%

Similar to mmWave communication, the high cost and the high power consumption in digital beamforming makes it unsuitable for THz communication. The analog beamforming at THz waveband can reduce the number of required phase shifters in the RF domain. Nevertheless, it is subject to the additional hardware constraints because the analog phase shifters are digitally controlled and just have quantized phase values, which will significantly restrict analog beamforming performance in practice. On the other hand, the hybrid analog/digital beamforming is again a better trade-off between the analog and digital methods. The hybrid beamforming can have fewer RF chains than antennas and approaches the fully digital performance in sparse channels \cite{Chen2019a}. 



The  types of antennas that can be used in THz communication are photoconductive antennas, horn antennas, lens antennas, microstrip antennas, and on-chip antennas.
Initially, THz antennas were designed inside the semiconductor using Indium Phosphide (InP) or Gallium Arsenide (GaAs) in which controlling radiation pattern was difficult due to high dielectric constant. Therefore lens-based antennas that were fed by horns were proposed. 
Other approaches, like stacking different substrate layers with different dielectric properties, were proposed to improve the antenna efficiency \cite{Jamshed2020}. 
 In addition to metallic antennas and dielectric antennas, antennas based on new materials are also possible {\em e.g.}, carbon-nanotube based antennas and the planar graphene antennas \cite{He2020}.

\subsection{Channel Models}  

In order to evaluate the performance of the communication system, the very first step is to construct an accurate channel model. Not surprisingly, researchers have developed different channel models for mmWave to be used in simulators and analysis. For example, in 2012, the mobile and wireless communications enablers for the twenty-twenty information society (METIS) project proposed three-channel models, namely the stochastic, the map-based and the hybrid model, where the stochastic model is suitable for frequencies up to $70$ GHz, while the map-based model is applicable for frequencies up to $100$ GHz. In 2017, 3GPP 3D channel model for the sub-100 GHz band was proposed. NYUSIM is another channel model developed with the help of real-world propagation channel measurements at mmWave frequencies ranging from $28$ GHz to $73$ GHz in different outdoor scenarios \cite{SSTS2017}. Statistical channel models for UM-MIMO are classified into \textit{matrix-based models} and \textit{reference-antenna-based models}. Matrix-based models characterize the properties of the complete channel transfer matrix. On the other hand, the reference-antenna-based models consider a reference transmitting and receiving antenna first and analyze the point to point propagation model between them. Then, based on this model, the complete channel matrix is statistically generated \cite{faisal2020ultramassive}. 

As discussed above already, THz channels exhibit very different propagation characteristics compared to the lower frequency bands. Therefore, modeling the channel and noise is essential for the accurate performance evaluation of the THz communications systems \cite{Tekbyk2019}. Even the free-space scenario is not straightforward to model in this case because of the significant level of molecular absorption. Therefore, one needs to be include an additional exponential term along with the power-law model in the path-loss equation. Overall, the peculiar propagation characteristics of THz waves already discussed in Section~\ref{sec:propagation} make their analysis challenging. 

Except for the recent measurements at the sub-THz frequencies \cite{Rappaport2019}, the rest of the THz channel modeling work is driven by ray tracing \cite{Han2015, Priebe2013, Moldovan2014} or statistical channel modeling \cite{THzChModel01,Priebe2013a,Kim2015,Kim2016,Guan2019, Elayan2017,Elayan2018,ref175}. In particular, a statistical model for THz channel based on a universal stochastic spatiotemporal model has been introduced in \cite{Priebe2013a} for indoor channels ranging from $275$ GHz to $325$ GHz. A 2D geometrical statistical model for device-to-device scatter channels at sub-terahertz (sub-THz) frequencies was proposed in \cite{Kim2015, Kim2016}. In addition to the outdoor channel model,  the indoor model for intra-wagon channel characterization at $300$ GHz was discussed in \cite{Guan2019}. At nano-scale level, the channel models were presented in \cite{Elayan2017,Elayan2018} for intra body THz communication. A hybrid channel model was discussed in \cite{ref175} for chip-to-chip communication via THz frequencies.
  

We now describe a simple yet powerful analytically tractable channel model which can be adapted to various propagation scenarios. This is suitable for the system-level performance analysis including using ideas from stochastic geometry.


\subsubsection{mmWave Channel}
Consider a link between a transmitter and a receiver located at $r$ distance apart of type $s$ where $s\in\{\L,\N\}$ denoting whether the link is LOS and NLOS \cite{JGA2017}. For simplicity, let us assume narrow-band communication and analog beamforming. The received power at the receiver is given as
\begin{align}
	P_\rcvd=P_\tx \ell_s(r) g_\rx(\theta_\rx) g_\tx(\theta_\tx) H \label{eq:mmwavechannel}
\end{align}
where 
\begin{enumerate}
\item $\ell_s(r)$ denotes the standard path loss at distance $r$ which is due to spreading loss. It is given by a path-loss function, typically modelled using power-law as
$$\ell_s(r)=c_s r^{-\alpha_s},$$
where $c_s$ is the near-field gain and $\alpha_s$ is the path-loss exponent.
\item $p_\tx$ is the transmit power,
\item $g_\tx$ and $g_\rx$ are the transmitter and receiver antenna patterns while $\theta_\tx$ and $\theta_\rx$ are the angles denoting beam-direction of the transmitter and receiver. Therefore, $g_\tx(\theta_\tx)$ and $g_\rx(\theta_\rx)$ are respectively the transmitter and receiver antenna gains.
\item $H$ denotes the small scale fading coefficient. Nakagami fading is often assumed with different parameters $\mu_\L$ and $\mu_N$ for LOS and NLOS link \cite{JGA2017}. Therefore $H$ is a Gamma random variable with parameter $\mu_s$.
\end{enumerate}

The above channel model can be extended for different environments and propagation scenarios, for example, to include multiple paths \cite{JGA2017}, multi-rank channel \cite{Heath2016, Kulkarni2016}, hybrid beam-forming \cite{Kulkarni2016} and massive MIMO. Since the specific bands with high absorption loss are avoided, the effect of molecular absorption can be ignored for mmWave communication. 


\subsubsection{THz Channel} 
Since atmospheric attenuation and scattering are prominent at THz frequencies, the THz channel model is expected to be different from the one discussed above for the mmWave communications. Due to the huge difference between LOS and NLOS links, most of the works have considered LOS links only \cite{Kokkoniemi2017, Kokkoniemi2018}. For simplicity, we will assume narrowband communication. If we consider a LOS link between a transmitter and a receiver located at $r$ distance apart of type $s$, the received power $ P_\rcvd$ is given by  \cite{Federici2010, Ma2018a}
\begin{align}
P_\rcvd=P_\tx \ell(r) g_\rx(\theta_\rx) g_\tx(\theta_\tx) \tau(r) \label{eq:thzchannel}
\end{align}
where $\tau(r)$ is an additional loss term due to molecular absorption defined in \eqref{eq:molabs}. In LOS links, path-loss can be given by free space path loss \ie
$$\ell(r)=\left(\frac{\lambda^2}{4\pi}\right)\frac1{4\pi r^2}.$$

The model can be extended to include scatters/reflectors. If $r_1$ is the distance between the transmitter
and the  surface while $r_2$ is the distance between the  surface and the receiver, then the scattered and reflected power are
\begin{align*}
	\mathrm{{P}_{\rcvd,S}}=P_\tx g_\rx(\theta_\rx) g_\tx(\theta_\tx) \ell(r_1)l(r_2)\tau(r_1)\tau(r_2) \Gamma_R
\end{align*}
and
\begin{align*}
	\mathrm{{P}_{\rcvd,R}}=P_\tx  g_\rx(\theta_\rx) g_\tx(\theta_\tx) \ell(r_1+r_2)\Gamma^2\tau(r_1+r_2)\Gamma_S,
\end{align*}
respectively, where $\Gamma_R$ and $\Gamma_S$ are coefficients related to reflection and scattering and may depend on surface orientation and properties.
The above channel model can be extended to include other scenarios, for example, multiple paths and wide-band communication \cite{Han2015}. 


\section{The mmWave Communications Systems} \label{sec:mmwave_comm}





As discussed above already, the major advantage of using mmWave communications is the availability of abundant spectrum, which is making multi-gigabit-per-second communication possible \cite{ZYP2011}. However, mmWave signals are more susceptible to blockages and foliage losses, which necessitates highly directional transmission. The combination of high signal attenuation and directional transmission offers several advantages and disadvantages for practical mmWave systems. On the positive side, these make mmWave systems more resilient to interference and hence more likely to operate in the noise-limited regime \cite{TS2013a}. Because of this, it is possible for the operators to use higher frequency reuse factor, thereby resulting in higher network capacity \cite{RS2018a,RS2018b}. For the same reasons, mmWave transmissions are inherently more secure compared to the sub-$6$ GHz transmissions \cite{NY2015,CW2016,YZ2017,QX2018}. For instance, the high attenuation of susceptibility to blockages make it difficult for the remote  eavesdroppers to even overhear mmWave transmissions unless they are located very close to the transmitters. Finally, as will be discussed in detail next, these reasons also make spectrum sharing more feasible at the mmWave frequencies. 

On the flip side, with high directivity, the initial cell search becomes a critical issue. Because of the use of directional beams, both the BSs and users need to perform a spatial search over a wide range of angles to align their transmission and reception beams in the correct direction. This adds significant delay and overhead to the communication. The situation degrades further when users are highly mobile due to increased occurrences of handovers. Further, higher susceptibility to blockages can result in outages. One approach to mitigate this is to utilize the concept of macro-diversity \cite{AKG2018b}, \cite{IKJ2019} and \cite{YZ2009}, where simultaneous connections with multiple BSs  are maintained for each user so that it does not experience any service interruption in the event of blocking of one BS. 

After summarizing these key features of mmWave communications, we now discuss a few key implications of these features on the system design. This section will be concluded with a discussion on the potential uses of mmWave communications in future 6G systems.

\subsection{Key System Design Implications}  


\subsubsection{Coexistence with lower frequency systems}  
Due to their limited transmission range, a mmWave system may not work effectively in a standalone deployment \cite{giordani2019standalone}. In particular, they need to coexist with conventional cellular networks operating on more favorable sub-$6$ GHz bands such that all the control level management, including load balancing and handovers, is performed over sub-$6$ GHz microwave transmissions while the data transmissions occur over the mmWave bands. Such networks will provide high capacity and better throughput in comparison to the standalone networks without decreasing reliability \cite{YN2015}. Further, macro-diversity can be utilized, where multiple BSs (some can be sub-6 GHz and some are mmWave) can connect to a user simultaneously to improve LOS probability and link throughput \cite{AKG2018b}. 


\subsubsection{Spectrum sharing} 
At lower frequencies, owning an {exclusive license} of a spectrum band ensures reliability and provides performance guarantees to applications with time-critical operations for an operator. However, mmWave systems often operate in a noise-limited regime because of which exclusive licensing at these frequencies may result in an under-utilization of the spectrum \cite{AG2020}. It has been shown that spectrum sharing at mmWave frequencies does not require sophisticated inter-cell coordination and even uncoordinated spectrum sharing between two or more operators is feasible \cite{GupAndHea2016}. This is an attractive option for the  {\em unlicensed spectrum} located at $59 - 64$ GHz and $64 - 71$ GHz bands which will allow multiple users to access the spectrum without any explicit coordination. Such a use of unlicensed spectrum increases spectrum utilization and helps minimize the entry barrier for new or small-scale operators. Even at licensed bands, shared use of spectrum can help increase the spectrum utilization and reduce licensing costs. It has also been shown that simple inter-cell interference coordination mechanisms can be used to improve the sharing performance \cite{AGAA2016, JGA2017, IH2018, SR2014,JGA2014}. Furthermore, the bands where mmWave communication coexists with other services (including incumbent services and newly deployed applications) may need to protect each other in the case of dense deployments. For this, spectrum license sharing mechanisms such as uncoordinated, static, and dynamic are the viable options in these bands. Spectrum sharing opportunities at mmWave bands also bring the need to evolve new methods of spectrum licensing which need to be more flexible, opportunistic, dynamic and area specific \cite{AG2020}.


\subsubsection{Ultra-dense networks} 
Ultra-dense networks (UDN) are characterized by very short inter-site distances. They are generally used to provide local coverage in highly populated residential areas, office buildings, university campuses, and city centers. The mmWave frequencies are a natural candidate for UDNs because of the directional transmission and blockage sensitivity which limits interference even at ultra dense deployments. Further, self backhauling provides an inexpensive way to connect these densely deployed APs/BSs to backhaul.



\subsubsection{Deep learning-based beamforming} 
The performance of the mmWave systems in a high mobility scenario is severely affected by large training overhead, which occurs due to the frequent updating of large array beamforming vectors. In the last few years, deep learning-based beamforming techniques have attracted considerable interest due to their ability to reduce this training overhead. At the transmitter, pilot signals from the UE are first transmitted to learn the RF signature of the neighboring environment and then this knowledge is used to predict the best beamforming vectors for the transmitted data RF signature. Thus, after successful learning phase, the deep learning models require negligible training overhead which ensures reliable coverage and low latency for the mmWave applications \cite{AA2018}.


\subsection{Potential Applications of mmWave Communications in 6G}

\subsubsection{Wireless access applications}
Due to the abundance of bandwidth around $60$ GHz band, various technologies are expected to be developed to support unlicensed operations for the wireless local and personal area networks (WLANSs and WPANs) with potential applications in internet access at home, offices, transportation centers, and city hotspots.  These technologies are expected to support multi-gigabit data transmission, with examples including IEEE 802.11ad and IEEE 802.11ay \cite{TN2014,YG2017}. In future, IEEE 802.11ay based mmWave distribution networks (mDNs) may become an alternative-low-cost solution for the fixed optical fiber links. The purpose of mDNs is to provide  point-to-point (P2P) and point-to-multi-point (P2MP) mmWave access in indoor as well as outdoor scenario as well as wireless backhaul services to the small cells in an ad-hoc network scenario. The benefits of IEEE 802.11ay based mDN networks are cheaper network infrastructure and the high-speed ubiquitous coverage, while the major challenges include dealing with blockages, interference management, and developing efficient beam-training algorithms \cite{YL2020,KA2020}. Also, 5G is seen as a significant step in enabling cellular communication over mmWave bands which is expected to mature further in 6G and beyond systems. 

\subsubsection{Backhaul infrastructure} 
It is well known that providing fiber backhaul in highly dense small cell deployments is challenging due to increased installation and operational cost~\cite{dhillon2015wireless,saha2018bandwidth,saha2019millimeter}. 
Not surprisingly, many researchers have recently invested their efforts to enable wireless backhaul in mmWave bands owing to their directional communication and high LOS throughput. The present 5G cellular backhaul networks are expected to operate on the $60$ GHz and $71 - 86$ GHz bands, which are expected to be extended to the $92 - 114.25$ GHz band due to its similar propagation characteristics. Significant efforts have also gone into developing new technologies including cross-polarization interference cancelation (XPIC), bands and carriers aggregation (BCA), LOS MIMO, orbital angular momentum (OAM) in order to increase the capacity of the current mmWave backhaul solutions \cite{LR2019}. Further work is needed to provide backhaul solutions for the data-hungry future applications of 6G by using higher mmWave bands (above $100$ GHz) as well as the THz spectrum \cite{MJ2016, LR2019}.

\subsubsection{Information showers} Information showers (ISs) are high bandwidth ultra-short range hot spots in which mmWave BSs operating at the unlicensed $60$ GHz band are mounted on the ceilings, doorways, entrances of the commercial buildings or pavements, which deliver multi-gigabit data rates over a coverage range of about  $10$ m \cite{TSR2015}. Thus they provide an ideal platform to exchange a huge amount of data  between different kind of networks, devices and users over a very short span of time. Unlike conventional small-cell cellular networks, ISs can be used for both offloading as well as pre-fetching of data from the long haul wireless network for applications like instant file transfer and video streaming. ISs also help in improving the energy efficiency and battery life of the mobile terminal due to its ability to download videos and large files within a few seconds. 
However, installations of ISs requires a very robust architecture that can work seamlessly with the current cellular networks and is still an open area of research \cite{SB2019,SJ2019}.

\subsubsection{Aerial communications} Many frequency bands in the mmWave spectrum region are already being used to support the high-capacity satellite to ground transmission. However, with the maturity of the drone technology, the future wireless networks is expected to have a much more dynamic aerial component with drones used in a diverse set of applications, such as agriculture, mapping, traffic control, photography, surveillance, package delivery, telemetry, and on-demand handling of higher network loads in large public gatherings like music concerts. Because of higher likelihood of LOS in many of these applications, mmWave communications is expected to play a particularly promising role. Further, quick (on-demand) and easy deployment of drones also make them attractive for many public safety applications, especially when the civil communication infrastructure is compromised or damaged. Naturally, mmWave communications can play a promising role in such applications as well.

\subsubsection{Vehicular communications}
The ability of vehicles to communicate among themselves as well as the wireless infrastructure not only helps in the navigation of completely autonomous vehicles but is also helpful in the avoidance of accidents in semi-autonomous and manually driven vehicles through timely alerts and route guidance~\cite{dhillon2020poisson}. Because of the high likelihood of LOS and the need to support high data rates, the mmWave (and THz) spectrum is naturally being considered for the vehicular communications systems \cite{PK2015,ben2011millimeter, JH2012,YH2016,VP2018}. Further, unified vehicular communications and radar sensing mechanisms are needed for the massive deployments of interconnected smart cars which can easily cope with the rapidly maturing automotive environments, consisting of the networked road signs, connected pedestrians, video surveillance systems, and smart transportation facilities \cite{VP2019}.

\section{The THz Communications Systems}

After discussing mmWave communications systems in detail in the previous section, we now focus on the THz communications systems in this section. Since the THz band is higher in frequency than the mmWave band, the communication at THz band faces almost all the critical challenges that we discussed in the context of mmWave communications. In order to avoid repetition, we will therefore focus on the challenges and implications that are more unique (or at least more pronounced) to the THz communications systems. 


\subsubsection{Smaller range} 
Due to  high propagation and molecular absorption losses, communication range of THz bands is further limited compared to the mmWave transmission. For instance, in small cells, the THz band may provide coverage up to only about $10$ m \cite{Akyildiz2014}. Further, the frequency-dependent molecular absorption in the THz bands results in band-splitting and bandwidth reduction \cite{Sarieddeen2020}. 


\subsubsection{THz transceiver design} In THz communication, the transceivers need to be wideband, which is a major challenge. The frequency band of the signal to be generated is too high for conventional oscillators, while it is too low for optical photon emitters. This problem is known as the \textit{THz gap}. Another challenge is the design of antennas and amplifiers which support ultra-wideband transmission for THz communication \cite{Tekbyk2019}. Currently, the THz waves are generated using either conventional oscillators or optical photon emitters along with frequency multiplier/divider. 


\subsubsection{THz beam tracking} 
Just like mmWave systems, the THz communications systems require beamforming to overcome large propagation losses. However, beamforming requires channel state information, which is challenging to obtain when the array sizes are large, as is the case in THz communications systems. Therefore, it is vital to accurately measure the AoD of transmitters and the angle of arrival (AoA) of receivers using beam tracking techniques. While such beam tracking techniques have been studied extensively for the lower frequencies, it is not so for the THz frequencies. In THz communication, in order to achieve beam alignment, beam switching must be done before beam tracking. However, due to large array sizes, the codebook design for beam switching is computationally complex. On the flip side, these complex codebooks will generate high-resolution beams, which help in accurate angle estimation \cite{Chen2019a}. This provides a concrete example of the type of challenges and subtle tradeoffs that need to be carefully understood while designing THz communications systems.




The implications of these challenges are similar to the ones we discussed for the mmWave systems in the previous section. For instance, due to the limited coverage area, the THz communication systems are more likely to be deployed for indoor applications \cite{Singh2019}. In particular, the indoor links have been found to be robust even in the presence of one or two NLOS reflection components \cite{Ma2018a}. Likewise, owing to small coverage areas and high directionality, it is expected that the THz systems would efficiently share spectrum without much coordination (similar to the mmWave systems). In bands where passive services like radio astronomy and satellite-based earth monitoring are already present, THz communication systems need to share the spectrum under some protection rules. 


\subsection{Potential Applications of THz Communications in 6G}

THz communication has many applications in macro as well as in micro/nano scale. Some of the applications are discussed in this section.



\subsubsection{Macroscale THz Communication} 

Most of the macroscale use cases of THz communications will be driven by emerging applications requiring Tbps links, which are not possible using mmWave spectrum. Such applications include ultra HD video conferencing and streaming, 3D gaming, extended reality, high-definition holographic video conferencing, haptic communications, and tactile internet, to name a few. Within conventional cellular network settings, the THz bands are most suitable for small cell indoor applications or high-speed wireless backhaul for small cells \cite{Akyildiz2014}. Likewise, in the conventional WLAN applications, the Terabit Wireless Local Area Networks (T-WLAN) can provide seamless interconnection between high-speed wired networks, such as optical fiber links, and personal devices, such as mobile phones, laptops, and smart TVs. Along similar lines, the Terabit Wireless Personal Area Networks (T-WPAN) can enable ultra-high-speed communication among proximate devices. A special type of WPAN application is {\em kiosk downloading}, where a fixed kiosk download station is used to transfer multimedia contents, such as large videos, to mobile phones located in its proximity \cite{Kurner2014}. Other potential applications and advantages of THz communications, such as enhanced security, relevance for aerial and vehicular communications \cite{Saeed2020,Rasheed2020}, as well as the potential use for providing wireless connections in data centers, can be argued along the same lines as we did already for the mmWave networks in Section~\ref{sec:mmwave_comm}. In order to avoid repetition, we do not discuss this again.

\subsubsection{Micro/nanoscale THz  Communication}
 The THz band can also be used for enabling communications between nanomachines \cite{Akyildiz2014}. These nanomachines can perform simple tasks, such as computations, data storage, actuation, and sensing. Depending on the application, the transmission distance can vary from a few micrometers to a few meters. Some representative applications of the nanomachine communications are discussed below.

\begin{enumerate}
\item[(i)] \textit{Health monitoring:} Nanosensors or nanomachines deployed inside the human body can measure the level of glucose, cholesterol, the concentration of various ions, biomarkers emitted by the cancer cells, etc. \cite{Akyildiz2014}. The measured data can be wirelessly transmitted to a device outside the human body (e.g., mobile phone or a smart band) using THz communication. The external device can process the data and further send it to a medical equipment or to a doctor. 

\item[(ii)] \textit{Nuclear, biological and chemical defenses:} Nanosensors are capable of sensing harmful chemical and bio-weapon molecules effectively \cite{Akyildiz2014}. In contrast to the classical macro-scale chemical sensors, the nanosensors can detect very small concentrations (as small as a single molecule). As a result, the nano devices communicating in the THz bands can be used in defense applications for the detection of harmful chemical, biological and nuclear agents.

\item[(iv)] \textit{Internet of nano-things (IoNT) and Internet of bio-nano-things (IoBNT):} The interconnection of nanomachines with the existing communication network is known as IoNT \cite{Akyildiz2010}. These interconnected nanodevices via IoNT can serve a variety of purposes ranging from tracking atmospheric conditions and health status to enabling real-time tracking. Also, nano-transceivers and antennas can be embedded in nearly all devices to be connected to the Internet. IoBNT is conceptually the same as IoNT but consists of biological nanomachines as opposed to the silicon-based nanomachines \cite{Akyildiz2015}. Biological nanomachines can be made from synthetic biological materials and a modified cell via genetic engineering. IoBNT has many applications in the biomedical field.

\item[(iv)] \textit{Wireless network on-chip communication: } THz waves can enable communication among processing cores embedded on-chip with the help of planar nano-antennas of a few micrometers in size \cite{Abadal2013}. This creates ultra-high-speed inter-core communication for applications where area is a constraint. Graphene based nano-antennas can be used for the design of scalable and flexible wireless networks on the chips. 

\end{enumerate}



\subsection{Nanonetworks}

While the capabilities of a single nanomachine are limited to simple computations, sensing, and actuation, a network of inter-connected nanomachines can perform much more complex tasks. The nanomachines can communicate with each other or with a central device. Such networks have a wide variety of applications ranging from cancer treatment to environmental monitoring. The two potential carriers of information between nanomachines are EM waves and chemical molecules. Inside the human body, molecular communication has several advantages over EM waves, such as bio-compatibility and energy efficiency. 

\subsubsubsection{Integration with molecular communication} A nano-scale communication network consist of five fundamental components \cite{Yang2020}: 

\begin{enumerate}
\item[(i)] \textit{Message carrier:} Chemical molecules or waves that carry information from the transmitter to the receiver.

\item[(ii)] \textit{Motion component:} Provides force that is needed for the message carrier to move in the communication medium.

\item[(iii)] \textit{Field component:} Guides the message carrier in the communication medium. External fields include the EM field, molecular motors, and non-turbulent fluid flow. Internal fields include swarm motion or flocking behavior.

\item[(iv)] \textit{Perturbation:} This represents the variation of the message carrier to represent the transmit information. Perturbation is similar to modulation in telecommunication. It can be achieved by varying the concentration or type of molecules based on the transmit information.

\item[(v)] \textit{Specificity:} Reception process of the message at the target. For example, the binding of molecules with the receptor structures present in the target. 
\end{enumerate}

\begin{figure}[ht!]
	\centering
	\includegraphics[width=.8\linewidth]{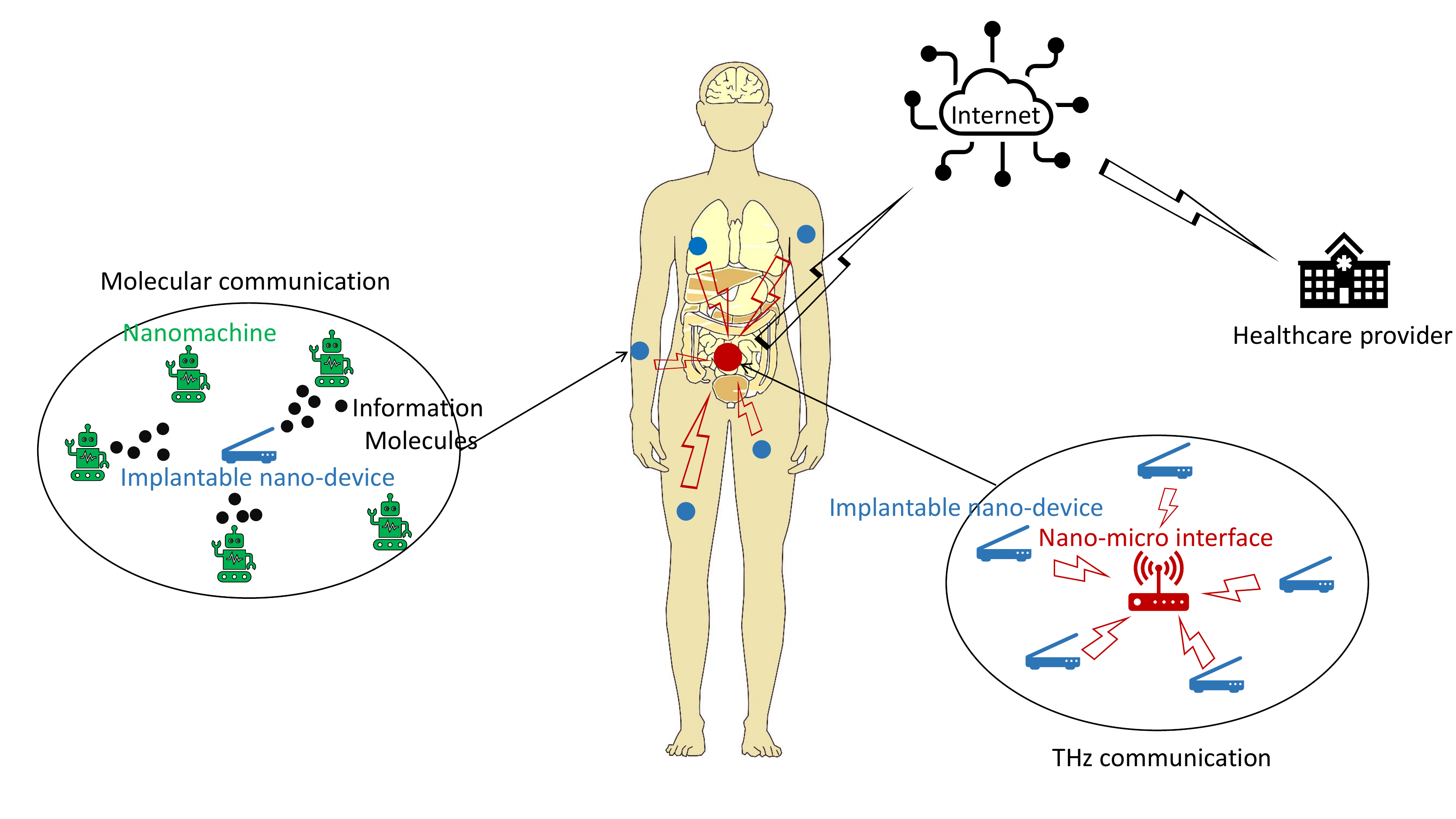}
	\caption{A hybrid nano communication network showing an eco-system consisting of the biological and artificial components, where various communication technologies including molecular and THz communication may co-exist together \cite{Yang2020}.}
	\label{fig:hybrid}
\end{figure}

 A hybrid communication system that combines molecular and EM paradigms was proposed in \cite{Yang2020} (See Fig. \ref{fig:hybrid}). In this hybrid communication network, MC is used inside the body due to its bio-compatibility, energy efficiency, and the lack of need for communication infrastructure for propagation methods like diffusion-based MC (molecules propagates in the medium based on concentration gradient). The nano-nodes (bio-nanomachines) form clusters and sense the data locally. A bio-nanomachine collects the health parameters, modulates the data, and transmits the information to the other bio-nanomachines (that act as relays). Now, for delivering gathered information to a receiver outside the human body, a graphene-based nano-device is implanted into the body. This implantable nano-device is made up of a chemical nanosensor, a transceiver, and the battery. Based on the concentration of information molecules transmitted by the bio-nanomachines to the implantable nano-device, the concentration is converted to a corresponding electrical signal. Now the implantable nano-devices communicate to the nano-micro interface via THz waves. This interface can be a dermal display or a micro-gateway to connect to the internet. {\em This type of hybrid communication network is bio-compatible due to MC technology and well connected to the outside world via THz communication.}


\section{Standardization Efforts}
We conclude this chapter by discussing the key standardization efforts for both the mmWave and THz communications systems. 

\subsection{Standardization Efforts for mmWave Communications} 

Given the increasing interest in mmWave communications over the past decade, it is not surprising to note that several industrial standards have been developed for its use. Some of these standards are discussed below. 

\subsubsubsection{IEEE 802.11ad} 
This standard is focused on enabling wireless communications in the $60$ GHz band. It specifies amendments to the 802.11 physical and MAC layers to support multi-gigabit wireless applications in the $60$ GHz band. The unlicensed spectrum around $60$ GHz has approximately $14$ GHz bandwidth, which is divided into channels of $2.16$, $4.32$, $6.48$, and $8.64$ GHz bandwidth. The IEEE 802.11ad standard supports transmission rates of up to $8$ Gbps using single-input-single-output (SISO) wireless transmissions over a single $2.16$ GHz channel. It supports backward compatibility with existing Wi-Fi standards in the $2.4$ and $5$ GHz bands. Therefore, future handsets may have three transceivers operating at $2.4$ GHz (for general use), $5$ GHz (for higher speed applications), and $60$ GHz (for ultra-high-speed data applications) \cite{TN2014,YN2015,NAF2019}. 

\subsubsubsection{IEEE 802.11ay} This standard is the enhancement of IEEE 802.11ad  to support fixed point-to-point (P2P) and point-to-multipoint (P2MP) ultra-high-speed indoor and outdoor mmWave communications. It supports channel bonding and aggregation to enable $100$ Gbps data rate. Channel bonding allows a single waveform to cover at least two or more contiguous $2.16$ GHz channels whereas channel aggregation allows a separate waveform for each aggregated channel \cite{YG2017,PZ2018}. 

\subsubsubsection{IEEE 802.15.3c} 
This standard defines the physical and MAC layers for the indoor $60$ GHz WPANs. In this standard, the MAC implements a random channel access and time division multiple access approaches to support the directional and the quasi-omnidirectional transmissions \cite{NAF2019}. 

\subsubsubsection{ECMA-387} 
This standard proposed by the European computer manufacturers association (ECMA) specifies the physical layer, the MAC layer and the high-definition multimedia interface (HDMI) protocol adaptation layer (PAL) for the $60$ GHz wireless networks. The ECMA-387 standards can be applied to the handheld devices used for low data rate transfer at short distances and to the devices equipped with adaptive antennas used for high data rate multimedia streaming at longer distances \cite{NAF2019}. 

\subsubsubsection{5G NR mmWave standard} 
This is a global standard platform for the 5G wireless air interfaces connecting mobile devices to the 5G base stations. Together with the efforts of IMT2020, the 3GPP Release 15 gave the first set of standards detailing 5G NR use cases, broadly categorized as the eMBB, the uRLLC and the mMTC. However, Release 15 is solely dedicated to the non-standalone (NSA) operation of 5G NR in which 4G LTE networks and 5G mobile technology co-exist \cite{JP2020}. The 3GPP Release 16 (completed in July 2020) targeted new enhancements for the better performance of standalone (SA) 5G NR networks (operating in the range of $1 - 52.6$ GHz) in terms of increased capacity, improved reliability, reduced latency, better coverage, easier deployment, power requirements, and mobility. It also includes discussions on multi-beam management, over the air (OTA) synchronization to support multi-hop backhauling, integrated access and backhaul (IAB) enhancements, remote interference management (RIM) reference signals, UE power savings and mobility enhancements  \cite{JP2020}. The next release is expected to be completed in September 2021, which will address the further enhancements for the 5G NR. It will include the support for new services like critical medical applications, NR broadcast, multicast and multi SIM devices, mission critical applications, cyber security applications, and dynamic spectrum sharing improvements, to name a few \cite{JP2020}.


\subsection{Standardization Efforts for THz Communications} 

Given that the THz communications is still in its nascent phase, its standardization efforts are just beginning. 
The IEEE 802.15.3d-2017 was proposed in 2017, which is the first standard for THz fixed point-to-point links operating at carrier frequencies between $252$ and $321$ GHz using eight different channel bandwidths ($2.16$ GHz to $69.12$ GHz and multiples of $2.16$ GHz). For the development of nano-network standards at THz frequencies, \textit{IEEE P1906.1/Draft 1.0} discusses recommended practices for nano-scale and molecular communication frameworks \cite{molstd,Elayan2020,Kurner2020}.

\section{Conclusion}

The main claim to fame for the 5G communications systems is to demonstrate that mmWave frequencies can be efficiently used for commercial wireless communications systems, which until only a few years back was considered unrealistic because of the unfavorable propagation characteristics of these frequencies. Even though the 5G systems are still being rolled out, it is argued that the gigabit-per-second rates to be supported by the 5G mmWave systems may fall short in supporting many emerging applications, such as 3D gaming and extended reality. Such applications will require several hundreds of gigabits per second to several terabits per second data rates with low latency and high reliability, which are currently considered to be the design goals for the next generation 6G communications systems. Given the potential of THz communications systems to provide such rates over short distances, they are currently considered to be the next frontier for wireless communications research. Given the importance of both mmWave and THz bands in 6G and beyond systems, this chapter has provided a unified treatment of these bands with particular emphasis on their propagation characteristics, channel models, design and implementation considerations, and potential applications. The chapter was concluded with a brief survey of the current standardization activities for these bands.


	


\bibliographystyle{IEEEtran}
\bibliography{thz_6g}

\begin{thebibliography}{100}
\providecommand{\url}[1]{#1}
\csname url@samestyle\endcsname
\providecommand{\newblock}{\relax}
\providecommand{\bibinfo}[2]{#2}
\providecommand{\BIBentrySTDinterwordspacing}{\spaceskip=0pt\relax}
\providecommand{\BIBentryALTinterwordstretchfactor}{4}
\providecommand{\BIBentryALTinterwordspacing}{\spaceskip=\fontdimen2\font plus
\BIBentryALTinterwordstretchfactor\fontdimen3\font minus
  \fontdimen4\font\relax}
\providecommand{\BIBforeignlanguage}[2]{{%
\expandafter\ifx\csname l@#1\endcsname\relax
\typeout{** WARNING: IEEEtran.bst: No hyphenation pattern has been}%
\typeout{** loaded for the language `#1'. Using the pattern for}%
\typeout{** the default language instead.}%
\else
\language=\csname l@#1\endcsname
\fi
#2}}
\providecommand{\BIBdecl}{\relax}
\BIBdecl

\bibitem{thebook}
Y.~Wu, S.~Singh, T.~Taleb, A.~Roy, H.~S. Dhillon, M.~R. Kanagarathinam, and
  A.~De, Eds., \emph{{6G} mobile wireless networks}.\hskip 1em plus 0.5em minus
  0.4em\relax Springer, 2021.

\bibitem{TS2013a}
T.~S. Rappaport, S.~Sun, R.~Mayzus, H.~Zhao, Y.~Azar, K.~Wang, G.~N. Wong,
  J.~K. Schulz, M.~Samimi, and F.~Gutierrez, ``{Millimeter wave mobile
  communications for 5G cellular: It will work!}'' \emph{IEEE access}, vol.~1,
  pp. 335--349, 2013.

\bibitem{Sarieddeen2020}
H.~Sarieddeen, N.~Saeed, T.~Y. Al-Naffouri, and M.-S. Alouini, ``{Next
  generation terahertz communications: A rendezvous of sensing, imaging, and
  localization},'' \emph{IEEE Communication Magzine}, vol.~58, no.~5, pp.
  69--75, May 2020.

\bibitem{FK2012}
F.~Khan, Z.~Pi, and S.~Rajagopal, ``{Millimeter-wave mobile broadband with
  large scale spatial processing for 5G mobile communication},'' in \emph{2012
  50th annual allerton conference on communication, control, and computing
  (Allerton)}.\hskip 1em plus 0.5em minus 0.4em\relax IEEE, 2012, pp.
  1517--1523.

\bibitem{KLL2020}
K.~L. Lueth, ``{IoT 2019 in review: The 10 most relevant IoT developments of
  the year},'' 2020.

\bibitem{IOT2015}
{International Telecommunication Union}, ``{IMT traffic estimates for the years
  2020 to 2030},'' \emph{Report ITU}, pp. 2370--0, 2015.

\bibitem{IOT2020}
N.~M. Karie, N.~M. Sahri, and P.~Haskell-Dowland, ``{IoT threat detection
  advances, challenges and future directions},'' in \emph{2020 Workshop on
  Emerging Technologies for Security in IoT}.\hskip 1em plus 0.5em minus
  0.4em\relax IEEE, 2020, pp. 22--29.

\bibitem{dhillon2017wide}
H.~S. Dhillon, H.~Huang, and H.~Viswanathan, ``Wide-area wireless communication
  challenges for the internet of things,'' \emph{IEEE Communications Magazine},
  vol.~55, no.~2, pp. 168--174, Feb. 2017.

\bibitem{AG2020}
A.~K. Gupta and A.~Banerjee, ``Spectrum above radio bands,'' \emph{{Spectrum
  Sharing: The Next Frontier in Wireless Networks}}, pp. 75--96, 2020.

\bibitem{JGA2014}
J.~G. Andrews, S.~Buzzi, W.~Choi, S.~V. Hanly, A.~Lozano, A.~C. Soong, and
  J.~C. Zhang, ``{What will 5G be?}'' \emph{IEEE Journal on Selected Areas in
  Communications}, vol.~32, no.~6, pp. 1065--1082, 2014.

\bibitem{GupSabDhi2020}
A.~K. Gupta, N.~V. Sabu, and H.~S. Dhillon, ``Fundamentals of network
  densification,'' in \emph{{5G} and Beyond: Fundamentals and Standards}.\hskip
  1em plus 0.5em minus 0.4em\relax Springer, 2020.

\bibitem{AndZDG2016}
J.~G. Andrews, X.~Zhang, G.~D. Durgin, and A.~K. Gupta, ``Are we approaching
  the fundamental limits of wireless network densification?'' \emph{IEEE
  Communication Magazine}, vol.~54, pp. 184--190, Oct. 2016.

\bibitem{Han2019}
C.~Han, Y.~Wu, Z.~Chen, and X.~Wang, ``{Terahertz Communications (TeraCom):
  Challenges and impact on 6G wireless systems},'' pp. 1--8, Dec. 2019.

\bibitem{saad2019vision}
W.~Saad, M.~Bennis, and M.~Chen, ``A vision of {6G} wireless systems:
  Applications, trends, technologies, and open research problems,'' \emph{IEEE
  network}, vol.~34, no.~3, pp. 134--142, 2019.

\bibitem{JGA2017}
J.~G. Andrews, T.~Bai, M.~N. Kulkarni, A.~Alkhateeb, A.~K. Gupta, and R.~W.
  Heath, ``{Modeling and analyzing millimeter wave cellular systems},''
  \emph{IEEE Transactions on Communications}, vol.~65, no.~1, pp. 403--430,
  2016.

\bibitem{Akyildiz2010}
I.~Akyildiz and J.~Jornet, ``{The internet of nano-things},'' \emph{IEEE
  Wireless Communication}, vol.~17, no.~6, pp. 58--63, Dec. 2010.

\bibitem{Yang2020}
K.~Yang, D.~Bi, Y.~Deng, R.~Zhang, M.~M.~U. Rahman, N.~A. Ali, M.~A. Imran,
  J.~M. Jornet, Q.~H. Abbasi, and A.~Alomainy, ``{A comprehensive survey on
  hybrid communication in context of molecular communication and terahertz
  communication for body-centric nanonetworks},'' \emph{IEEE Transaction on
  Molecular, Biological and Multi-Scale Communication}, pp. 1--1, 2020.

\bibitem{Sabu2019}
N.~V. Sabu and A.~K. Gupta, ``{Analysis of diffusion based molecular
  communication with multiple transmitters having individual random information
  bits},'' \emph{IEEE Transaction on Molecular, Biological and Multi-Scale
  Communication}, vol.~5, no.~3, pp. 176--188, Dec. 2019.

\bibitem{ITU2019}
J.~ITU, ``{Provisional final acts},'' in \emph{World Radiocommunication
  Conference 2019}.\hskip 1em plus 0.5em minus 0.4em\relax ITU Publications,
  2019.

\bibitem{Kurner2020}
T.~Kurner and A.~Hirata, ``{On the impact of the results of WRC 2019 on THz
  communications},'' in \emph{Third International Workshop on Mobile Terahertz
  System}.\hskip 1em plus 0.5em minus 0.4em\relax IEEE, July 2020, pp. 1--3.

\bibitem{ITU2015}
P.~F. Acts, ``{World radiocommunication conference (WRC-15)}.''\hskip 1em plus
  0.5em minus 0.4em\relax ITU, 2015.

\bibitem{ED2020}
E.~Dahlman, S.~Parkvall, and J.~Skold, \emph{{5G NR: The next generation
  wireless access technology}}.\hskip 1em plus 0.5em minus 0.4em\relax Academic
  Press, 2020.

\bibitem{FB2014}
F.~Boccardi, R.~W. Heath, A.~Lozano, T.~L. Marzetta, and P.~Popovski, ``{Five
  disruptive technology directions for 5G},'' \emph{IEEE Communications
  Magazine}, vol.~52, no.~2, pp. 74--80, 2014.

\bibitem{IH2018}
I.~A. Hemadeh, K.~Satyanarayana, M.~El-Hajjar, and L.~Hanzo, ``{Millimeter-wave
  communications: Physical channel models, design considerations, antenna
  constructions, and link-budget},'' \emph{IEEE Communications Surveys and
  Tutorials}, vol.~20, no.~2, pp. 870--913, 2017.

\bibitem{SR2014}
S.~Rangan, T.~S. Rappaport, and E.~Erkip, ``{Millimeter-wave cellular wireless
  networks: Potentials and challenges},'' \emph{Proceedings of the IEEE}, vol.
  102, no.~3, pp. 366--385, 2014.

\bibitem{Petrov2017}
V.~Petrov, M.~Komarov, D.~Moltchanov, J.~M. Jornet, and Y.~Koucheryavy,
  ``{Interference and SINR in millimeter wave and terahertz communication
  systems with blocking and directional antennas},'' \emph{IEEE Transaction
  Wireless Communication}, vol.~16, no.~3, pp. 1791--1808, Mar. 2017.

\bibitem{MLA2020}
M.~L. Attiah, A.~A.~M. Isa, Z.~Zakaria, M.~Abdulhameed, M.~K. Mohsen, and
  I.~Ali, ``{A survey of mmWave user association mechanisms and spectrum
  sharing approaches: An overview, open issues and challenges, future research
  trends},'' \emph{Journal on Wireless Networks}, vol.~26, no.~4, pp.
  2487--2514, 2020.

\bibitem{MM2005}
M.~Marcus and B.~Pattan, ``{Millimeter wave propagation: Spectrum management
  implications},'' \emph{IEEE Microwave Magazine}, vol.~6, no.~2, pp. 54--62,
  2005.

\bibitem{TripGupMeasurementFile2020}
\BIBentryALTinterwordspacing
S.~Tripathi and A.~Gupta, ``{Measurement efforts at mmWave indoor and outdoor
  environments}.'' [Online]. Available:
  \url{https://home.iitk.ac.in/~gkrabhi/memmwave}
\BIBentrySTDinterwordspacing

\bibitem{ART1988}
A.~Tharek and J.~McGeehan, ``{Propagation and bit error rate measurements
  within buildings in the millimeter wave band about 60 GHz},'' in \emph{8th
  European Conference on Electrotechnics, Conference Proceedings on Area
  Communication}.\hskip 1em plus 0.5em minus 0.4em\relax IEEE, 1988, pp.
  318--321.

\bibitem{allen199060}
G.~Allen and A.~Hammoudeh, ``{60 GHz propagation measurements within a
  building},'' in \emph{1990 20th European Microwave Conference}, vol.~2.\hskip
  1em plus 0.5em minus 0.4em\relax IEEE, 1990, pp. 1431--1436.

\bibitem{allen1991frequency}
------, ``{Frequency diversity propagation measurements for an indoor 60 GHz
  mobile radio link},'' in \emph{1991 Seventh International Conference on
  Antennas and Propagation}.\hskip 1em plus 0.5em minus 0.4em\relax IET, 1991,
  pp. 298--301.

\bibitem{davies1991wireless}
R.~Davies, M.~Bensebti, M.~Beach, and J.~McGeehan, ``{Wireless propagation
  measurements in indoor multipath environments at 1.7 GHz and 60 GHz for small
  cell systems},'' in \emph{[1991 Proceedings] 41st IEEE Vehicular Technology
  Conference}.\hskip 1em plus 0.5em minus 0.4em\relax IEEE, 1991, pp. 589--593.

\bibitem{yang2005impact}
H.~Yang, M.~H. Herben, and P.~F. Smulders, ``{Impact of antenna pattern and
  reflective environment on 60 GHz indoor radio channel characteristics},''
  \emph{IEEE Antennas and Wireless Propagation letters}, vol.~4, pp. 300--303,
  2005.

\bibitem{allen1991outdoor}
G.~Allen and A.~Hammoudeh, ``{Outdoor narrow band characterisation of
  millimetre wave mobile radio signals},'' in \emph{IEEE Colloquium on
  Radiocommunications in the Range 30-60 {GHz}}.\hskip 1em plus 0.5em minus
  0.4em\relax IET, 1991, pp. 4--1.

\bibitem{daniele1994outdoor}
N.~Daniele, D.~Chagnot, and C.~Fort, ``{Outdoor millimetre-wave propagation
  measurements with line of sight obstructed by natural elements},''
  \emph{Electronics Letters}, vol.~30, no.~18, pp. 1533--1534, 1994.

\bibitem{wang2005enhanced}
F.~Wang and K.~Sarabandi, ``{An enhanced millimeter-wave foliage propagation
  model},'' \emph{IEEE Transactions on Antennas and Propagation}, vol.~53,
  no.~7, pp. 2138--2145, 2005.

\bibitem{fong2005measurement}
B.~Fong, A.~Fong, G.~Hong, and H.~Ryu, ``{Measurement of attenuation and phase
  on 26-GHz wide-band point-to-multipoint signals under the influence of
  rain},'' \emph{IEEE Antennas and Wireless Propagation Letters}, vol.~4, pp.
  20--21, 2005.

\bibitem{ben2011millimeter}
E.~Ben-Dor, T.~S. Rappaport, Y.~Qiao, and S.~J. Lauffenburger,
  ``{Millimeter-wave 60 GHz outdoor and vehicle AOA propagation measurements
  using a broadband channel sounder},'' in \emph{2011 IEEE Global
  Telecommunications Conference-GLOBECOM 2011}.\hskip 1em plus 0.5em minus
  0.4em\relax IEEE, 2011, pp. 1--6.

\bibitem{rappaport2012broadband}
T.~S. Rappaport, F.~Gutierrez, E.~Ben-Dor, J.~N. Murdock, Y.~Qiao, and J.~I.
  Tamir, ``{Broadband millimeter-wave propagation measurements and models using
  adaptive-beam antennas for outdoor urban cellular communications},''
  \emph{IEEE Transactions on Antennas and Propagation}, vol.~61, no.~4, pp.
  1850--1859, 2012.

\bibitem{hirata2012}
A.~{Hirata}, H.~{Takahashi}, J.~{Takeuchi}, N.~{Kukutsu}, D.~{Kim}, and
  J.~{Hirokawa}, ``{120-GHz-band antenna technologies for over-10-Gbps wireless
  data transmission},'' in \emph{2012 6th European Conference on Antennas and
  Propagation (EUCAP)}, 2012, pp. 2564--2568.

\bibitem{keusgen2014propagation}
W.~Keusgen, R.~J. Weiler, M.~Peter, M.~Wisotzki, and B.~G{\"o}ktepe,
  ``{Propagation measurements and simulations for millimeter-wave mobile access
  in a busy urban environment},'' in \emph{2014 39th International Conference
  on Infrared, Millimeter, and Terahertz waves (IRMMW-THz)}.\hskip 1em plus
  0.5em minus 0.4em\relax IEEE, 2014, pp. 1--3.

\bibitem{weiler2014measuring}
R.~J. Weiler, M.~Peter, W.~Keusgen, and M.~Wisotzki, ``{Measuring the busy
  urban 60 GHz outdoor access radio channel},'' in \emph{2014 IEEE
  International Conference on Ultra-WideBand (ICUWB)}.\hskip 1em plus 0.5em
  minus 0.4em\relax IEEE, 2014, pp. 166--170.

\bibitem{guan2018towards1}
K.~Guan, B.~Ai, B.~Peng, D.~He, G.~Li, J.~Yang, Z.~Zhong, and T.~K{\"u}rner,
  ``{Towards realistic high-speed train channels at 5G millimeter-wave band -
  part I: Paradigm, significance analysis, and scenario reconstruction},''
  \emph{IEEE Transactions on Vehicular Technology}, vol.~67, no.~10, pp.
  9112--9128, 2018.

\bibitem{bas2019outdoor}
C.~U. Bas, R.~Wang, S.~Sangodoyin, T.~Choi, S.~Hur, K.~Whang, J.~Park, C.~J.
  Zhang, and A.~F. Molisch, ``{Outdoor to indoor propagation channel
  measurements at 28 GHz},'' \emph{IEEE Transactions on Wireless
  Communications}, vol.~18, no.~3, pp. 1477--1489, 2019.

\bibitem{Ma2018a}
J.~Ma, R.~Shrestha, L.~Moeller, and D.~M. Mittleman, ``{Invited article:
  Channel performance for indoor and outdoor terahertz wireless links},''
  \emph{APL Photonics}, vol.~3, no.~5, pp. 1--12, May 2018.

\bibitem{Federici2016}
J.~F. Federici, J.~Ma, and L.~Moeller, ``{Review of weather impact on outdoor
  terahertz wireless communication links},'' \emph{Nano Communication
  Networks}, vol.~10, pp. 13--26, Dec. 2016.

\bibitem{Priebe2012}
S.~Priebe, D.~M. Britz, M.~Jacob, S.~Sarkozy, K.~M. K.~H. Leong, J.~E. Logan,
  B.~S. Gorospe, and T.~Kurner, ``{Interference investigations of active
  communications and passive earth exploration services in the THz frequency
  range},'' \emph{IEEE Transaction on Terahertz Science Technology}, vol.~2,
  no.~5, pp. 525--537, Sep. 2012.

\bibitem{B.Heile}
B.~Heile, ``{ITU-R liaison request RE: active services in the band above 275
  {GHz}, IEEE standard 802.15-14-439-00-0THz},'' 2015.

\bibitem{Guan2019}
K.~Guan, B.~Peng, D.~He, J.~M. Eckhardt, S.~Rey, B.~Ai, Z.~Zhong, and
  T.~Kurner, ``{Channel characterization for intra-wagon communication at 60
  and 300 {GHz} bands},'' \emph{IEEE Transaction on Vehicular Technology},
  vol.~68, no.~6, pp. 5193--5207, June 2019.

\bibitem{Jornet2011}
J.~M. Jornet and I.~F. Akyildiz, ``{Channel modeling and capacity analysis for
  electromagnetic wireless nanonetworks in the terahertz band},'' \emph{IEEE
  Transaction on Wireless Communication}, vol.~10, no.~10, pp. 3211--3221, Oct.
  2011.

\bibitem{Rappaport2019}
T.~S. Rappaport, Y.~Xing, O.~Kanhere, S.~Ju, A.~Madanayake, S.~Mandal,
  A.~Alkhateeb, and G.~C. Trichopoulos, ``{Wireless communications and
  applications above 100 {GHz}: Opportunities and challenges for 6G and
  beyond},'' \emph{IEEE Access}, vol.~7, pp. 78\,729--78\,757, June 2019.

\bibitem{Kokkoniemi2015}
J.~Kokkoniemi, J.~Lehtom{\"{a}}ki, and M.~Juntti, ``{A discussion on molecular
  absorption noise in the terahertz band},'' \emph{Nano Communication
  Networks}, vol.~8, pp. 35--45, June 2016.

\bibitem{Lettington2002}
A.~H. Lettington, I.~M. Blankson, M.~F. Attia, and D.~Dunn, ``{Review of
  imaging architecture},'' \emph{Infrared Passive Millimeter-wave Imaging
  System Design, Analysis, Modelling and Testing}, vol. 4719, p. 327, 2002.

\bibitem{haneda20165g}
K.~Haneda, J.~Zhang, L.~Tan, G.~Liu, Y.~Zheng, H.~Asplund, J.~Li, Y.~Wang,
  D.~Steer, C.~Li \emph{et~al.}, ``{5G 3GPP-like channel models for outdoor
  urban microcellular and macrocellular environments},'' in \emph{2016 IEEE
  83rd Vehicular Technology Conference (VTC Spring)}.\hskip 1em plus 0.5em
  minus 0.4em\relax IEEE, 2016, pp. 1--7.

\bibitem{samimi2015probabilistic}
M.~K. Samimi, T.~S. Rappaport, and G.~R. MacCartney, ``{Probabilistic
  omnidirectional path loss models for millimeter-wave outdoor
  communications},'' \emph{IEEE Wireless Communications Letters}, vol.~4,
  no.~4, pp. 357--360, 2015.

\bibitem{bai2014analysis}
T.~Bai, R.~Vaze, and R.~W. Heath, ``{Analysis of blockage effects on urban
  cellular networks},'' \emph{IEEE Transactions on Wireless Communications},
  vol.~13, no.~9, pp. 5070--5083, 2014.

\bibitem{TB2014b}
T.~Bai and R.~W. Heath, ``{Coverage and rate analysis for millimeter-wave
  cellular networks},'' \emph{IEEE Transactions on Wireless Communications},
  vol.~14, no.~2, pp. 1100--1114, 2014.

\bibitem{gapeyenko2016analysis}
M.~Gapeyenko, A.~Samuylov, M.~Gerasimenko, D.~Moltchanov, S.~Singh, E.~Aryafar,
  S.~Yeh, N.~Himayat, S.~Andreev, and Y.~Koucheryavy, ``{Analysis of human-body
  blockage in urban millimeter-wave cellular communications},'' in \emph{2016
  IEEE International Conference on Communications (ICC)}.\hskip 1em plus 0.5em
  minus 0.4em\relax IEEE, 2016, pp. 1--7.

\bibitem{KV2016}
K.~Venugopal and R.~W. Heath, ``{Millimeter wave networked wearables in dense
  indoor environments},'' \emph{IEEE Access}, vol.~4, pp. 1205--1221, 2016.

\bibitem{bai2014}
T.~Bai and R.~W. Heath, ``{Analysis of self-body blocking effects in millimeter
  wave cellular networks},'' in \emph{2014 48th Asilomar Conference on Signals,
  Systems and Computers}.\hskip 1em plus 0.5em minus 0.4em\relax IEEE, 2014,
  pp. 1921--1925.

\bibitem{Ju2019}
S.~Ju, S.~H.~A. Shah, M.~A. Javed, J.~Li, G.~Palteru, J.~Robin, Y.~Xing,
  O.~Kanhere, and T.~S. Rappaport, ``{Scattering mechanisms and modeling for
  terahertz wireless communications},'' in \emph{Proceedings of ICC}.\hskip 1em
  plus 0.5em minus 0.4em\relax IEEE, May 2019, pp. 1--7.

\bibitem{Jarvelainen2012}
J.~Jarvelainen, K.~Haneda, M.~Kyro, V.-M. Kolmonen, J.-i. Takada, and
  H.~Hagiwara, ``{60 {GHz} radio wave propagation prediction in a hospital
  environment using an accurate room structural model},'' in \emph{2012
  Loughbrgh. Antennas and Propagation Conference}.\hskip 1em plus 0.5em minus
  0.4em\relax IEEE, Nov. 2012, pp. 1--4.

\bibitem{Degli-Esposti2007}
V.~Degli-Esposti, F.~Fuschini, E.~M. Vitucci, and G.~Falciasecca,
  ``{Measurement and modelling of scattering from buildings},'' \emph{IEEE
  Transaction on Antennas and Propagation}, vol.~55, no.~1, pp. 143--153, Jan.
  2007.

\bibitem{KulCorrectionFactor2018}
M.~N. {Kulkarni}, E.~{Visotsky}, and J.~G. {Andrews}, ``{Correction factor for
  analysis of MIMO wireless networks with highly directional beamforming},''
  \emph{IEEE Wireless Communications Letters}, vol.~7, no.~5, pp. 756--759,
  Oct. 2018.

\bibitem{Kokkoniemi2016}
J.~Kokkoniemi, P.~Rintanen, J.~Lehtomaki, and M.~Juntti, ``{Diffraction effects
  in terahertz band - Measurements and analysis},'' in \emph{Proceedings in
  GLOBECOM}.\hskip 1em plus 0.5em minus 0.4em\relax IEEE, Dec. 2016, pp. 1--6.

\bibitem{Federici2010}
J.~Federici and L.~Moeller, ``{Review of terahertz and subterahertz wireless
  communications},'' \emph{Journal of Applied Physics}, vol. 107, no.~11, pp.
  1--23, June 2010.

\bibitem{Bao2012}
L.~Bao, H.~Zhao, G.~Zheng, and X.~Ren, ``{Scintillation of THz transmission by
  atmospheric turbulence near the ground},'' in \emph{Fifth International
  Conference on Advanced Computational Intelligence}.\hskip 1em plus 0.5em
  minus 0.4em\relax IEEE, Oct. 2012, pp. 932--936.

\bibitem{CAB2016}
C.~A. Balanis, \emph{{Antenna theory: Analysis and design}}.\hskip 1em plus
  0.5em minus 0.4em\relax John wiley \& sons, 2016.

\bibitem{XY2017}
X.~Yu, J.~Zhang, M.~Haenggi, and K.~B. Letaief, ``{Coverage analysis for
  millimeter wave networks: The impact of directional antenna arrays},''
  \emph{IEEE Journal on Selected Areas in Communications}, vol.~35, no.~7, pp.
  1498--1512, 2017.

\bibitem{WL2015}
W.~Lu and M.~Di~Renzo, ``{Stochastic geometry modeling of cellular networks:
  Analysis, simulation and experimental validation},'' in \emph{Proceedings of
  the 18th ACM International Conference on Modeling, Analysis and Simulation of
  Wireless and Mobile Systems}, 2015, pp. 179--188.

\bibitem{MDR2016}
M.~Di~Renzo, W.~Lu, and P.~Guan, ``{The intensity matching approach: A
  tractable stochastic geometry approximation to system-level analysis of
  cellular networks},'' \emph{IEEE Transactions on Wireless Communications},
  vol.~15, no.~9, pp. 5963--5983, 2016.

\bibitem{AM2014}
A.~Maltsev, A.~Pudeyev, I.~Bolotin, G.~Morozov, I.~Karls, M.~Faerber, I.~Siaud,
  A.~Ulmer-Moll, J.~Conrat, R.~Weiler \emph{et~al.}, ``{MiWEBA D5.1: Channel
  modeling and characterization},'' \emph{Tech. Rep.}, 2014.

\bibitem{AT2015}
A.~Thornburg and R.~W. Heath, ``{Ergodic capacity in mmWave Ad Hoc network with
  imperfect beam alignment},'' in \emph{MILCOM 2015-2015 IEEE Military
  Communications Conference}.\hskip 1em plus 0.5em minus 0.4em\relax IEEE,
  2015, pp. 1479--1484.

\bibitem{DN2018}
N.~Deng and M.~Haenggi, ``{A novel approximate antenna pattern for directional
  antenna arrays},'' \emph{IEEE Wireless Communications Letters}, vol.~7,
  no.~5, pp. 832--835, 2018.

\bibitem{Chen2019a}
Z.~Chen, X.~Ma, B.~Zhang, Y.~Zhang, Z.~Niu, N.~Kuang, W.~Chen, L.~Li, and
  S.~Li, ``{A survey on terahertz communications},'' \emph{China
  Communication}, Feb. 2019.

\bibitem{Jamshed2020}
M.~A. Jamshed, A.~Nauman, M.~A.~B. Abbasi, and S.~W. Kim, ``{Antenna selection
  and designing for THz applications: Suitability and performance evaluation: A
  survey},'' \emph{IEEE Access}, vol.~8, pp. 113\,246--113\,261, 2020.

\bibitem{He2020}
Y.~He, Y.~Chen, L.~Zhang, S.-W. Wong, and Z.~N. Chen, ``{An overview of
  terahertz antennas},'' \emph{China Communication}, vol.~17, no.~7, pp.
  124--165, July 2020.

\bibitem{SSTS2017}
S.~Sun, G.~R. MacCartney, and T.~S. Rappaport, ``{A novel millimeter-wave
  channel simulator and applications for 5G wireless communications},'' in
  \emph{2017 IEEE International Conference on Communications (ICC)}.\hskip 1em
  plus 0.5em minus 0.4em\relax IEEE, 2017, pp. 1--7.

\bibitem{faisal2020ultramassive}
A.~Faisal, H.~Sarieddeen, H.~Dahrouj, T.~Y. Al-Naffouri, and M.-S. Alouini,
  ``{Ultramassive MIMO systems at terahertz bands: Prospects and challenges},''
  \emph{IEEE Vehicular Technology Magazine}, vol.~15, no.~4, pp. 33--42, Dec.
  2020.

\bibitem{Tekbyk2019}
K.~Tekbiyik, A.~R. Ekti, G.~K. Kurt, and A.~Gorcin, ``{Terahertz band
  communication systems: Challenges, novelties and standardization efforts},''
  \emph{Journal of Physics Communication}, vol.~35, pp. 53--62, Aug. 2019.

\bibitem{Han2015}
C.~Han, A.~O. Bicen, and I.~F. Akyildiz, ``{Multi-ray channel modeling and
  wideband characterization for wireless communications in the terahertz
  band},'' \emph{IEEE Transaction on Wireless Communication}, vol.~14, no.~5,
  pp. 2402--2412, May 2015.

\bibitem{Priebe2013}
S.~Priebe, M.~Kannicht, M.~Jacob, and T.~Kurner, ``{Ultra broadband indoor
  channel measurements and calibrated ray tracing propagation modeling at THz
  frequencies},'' \emph{Journal on Communication Networks}, vol.~15, no.~6, pp.
  547--558, Dec. 2013.

\bibitem{Moldovan2014}
A.~Moldovan, M.~A. Ruder, I.~F. Akyildiz, and W.~H. Gerstacker, ``{LOS and NLOS
  channel modeling for terahertz wireless communication with scattered rays},''
  in \emph{IEEE GLOBECOM Workshop}.\hskip 1em plus 0.5em minus 0.4em\relax
  IEEE, Dec. 2014, pp. 388--392.

\bibitem{THzChModel01}
\BIBentryALTinterwordspacing
Z.~Hossain, C.~Mollica, and J.~M. Jornet, ``Stochastic multipath channel
  modeling and power delay profile analysis for terahertz-band communication,''
  in \emph{Proceedings of the 4th ACM International Conference on Nanoscale
  Computing and Communication}, ser. NanoCom '17.\hskip 1em plus 0.5em minus
  0.4em\relax New York, NY, USA: Association for Computing Machinery, 2017.
  [Online]. Available: \url{https://doi.org/10.1145/3109453.3109473}
\BIBentrySTDinterwordspacing

\bibitem{Priebe2013a}
S.~Priebe and T.~Kurner, ``{Stochastic modeling of THz indoor radio
  channels},'' \emph{IEEE Transaction on Wireless Communication}, vol.~12,
  no.~9, pp. 4445--4455, Sep. 2013.

\bibitem{Kim2015}
S.~Kim and A.~Zajic, ``{Statistical modeling of THz scatter channels},'' in
  \emph{Ninth European Conference on Antennas and Propagation}, 2015.

\bibitem{Kim2016}
------, ``{Statistical modeling and simulation of short-range device-to-device
  communication channels at sub-THz frequencies},'' \emph{IEEE Transaction on
  Wireless Communication}, vol.~15, no.~9, pp. 6423--6433, Sep. 2016.

\bibitem{Elayan2017}
H.~Elayan, R.~M. Shubair, J.~M. Jornet, and P.~Johari, ``{Terahertz channel
  model and link budget analysis for intrabody nanoscale communication},''
  \emph{IEEE Transaction on Nanobioscience}, vol.~16, no.~6, pp. 491--503, Sep.
  2017.

\bibitem{Elayan2018}
H.~{Elayan}, C.~{Stefanini}, R.~M. {Shubair}, and J.~M. {Jornet}, ``End-to-end
  noise model for intra-body terahertz nanoscale communication,'' \emph{IEEE
  Transactions on NanoBioscience}, vol.~17, no.~4, pp. 464--473, 2018.

\bibitem{ref175}
C.~H. {Y. Chen}, ``{Channel modeling and analysis for wireless networks-on-chip
  communications in the millimeter wave and terahertz bands},''
  \emph{Proceedings of INFOCOM}, pp. 651--656, July 2018.

\bibitem{Heath2016}
R.~W. Heath, N.~Gonzalez-Prelcic, S.~Rangan, W.~Roh, and A.~M. Sayeed, ``{An
  overview of signal processing techniques for millimeter wave MIMO systems},''
  \emph{IEEE Journal on Selected Topics of Signal Processing}, vol.~10, no.~3,
  pp. 436--453, Apr. 2016.

\bibitem{Kulkarni2016}
M.~N. Kulkarni, A.~Ghosh, and J.~G. Andrews, ``{A comparison of MIMO techniques
  in downlink millimeter wave cellular networks with hybrid beamforming},''
  \emph{IEEE Transaction on Communication}, vol.~64, no.~5, pp. 1952--1967,
  May. 2016.

\bibitem{Kokkoniemi2017}
J.~Kokkoniemi, J.~Lehtomaki, and M.~Juntti, ``{Stochastic geometry analysis for
  mean interference power and outage probability in THz networks},'' \emph{IEEE
  Transaction on Wireless Communication}, vol.~16, no.~5, pp. 3017--3028, May
  2017.

\bibitem{Kokkoniemi2018}
J.~Kokkoniemi, J.~Lehtomaeki, and M.~Juntti, ``{Stochastic geometry analysis
  for band-limited terahertz band communications},'' \emph{IEEE Vehicular
  Technology Conference}, pp. 1--5, 2018.

\bibitem{ZYP2011}
Z.~Pi and F.~Khan, ``{An introduction to millimeter-wave mobile broadband
  systems},'' \emph{IEEE Communications Magazine}, vol.~49, no.~6, pp.
  101--107, 2011.

\bibitem{RS2018a}
R.~Sun, P.~B. Papazian, J.~Senic, C.~Gentile, and K.~A. Remley, ``{Angle- and
  delay-dispersion characteristics in a hallway and lobby at 60 GHz},'' 2018.

\bibitem{RS2018b}
R.~Sun, C.~A. Gentile, J.~Senic, P.~Vouras, P.~B. Papazian, N.~T. Golmie, and
  K.~A. Remley, ``{Millimeter-wave radio channels vs. synthetic beamwidth},''
  \emph{IEEE Communications Magazine}, vol.~56, no.~12, pp. 53--59, 2018.

\bibitem{NY2015}
N.~Yang, L.~Wang, G.~Geraci, M.~Elkashlan, J.~Yuan, and M.~Di~Renzo,
  ``{Safeguarding 5G wireless communication networks using physical layer
  security},'' \emph{IEEE Communications Magazine}, vol.~53, no.~4, pp. 20--27,
  2015.

\bibitem{CW2016}
C.~Wang and H.-M. Wang, ``{Physical layer security in millimeter wave cellular
  networks},'' \emph{IEEE Transactions on Wireless Communications}, vol.~15,
  no.~8, pp. 5569--5585, 2016.

\bibitem{YZ2017}
Y.~Zhu, L.~Wang, K.-K. Wong, and R.~W. Heath, ``{Secure communications in
  millimeter wave Ad Hoc networks},'' \emph{IEEE Transactions on Wireless
  Communications}, vol.~16, no.~5, pp. 3205--3217, 2017.

\bibitem{QX2018}
Q.~Xue, P.~Zhou, X.~Fang, and M.~Xiao, ``{Performance analysis of interference
  and eavesdropping immunity in narrow beam mmWave networks},'' \emph{IEEE
  Access}, vol.~6, pp. 67\,611--67\,624, 2018.

\bibitem{AKG2018b}
A.~K. Gupta, J.~G. Andrews, and R.~W. Heath, ``Macrodiversity in cellular
  networks with random blockages,'' \emph{IEEE Transactions on Wireless
  Communications}, vol.~17, no.~2, pp. 996--1010, Feb. 2017.

\bibitem{IKJ2019}
I.~K. Jain, R.~Kumar, and S.~S. Panwar, ``{The impact of mobile blockers on
  millimeter wave cellular systems},'' \emph{IEEE Journal on Selected Areas in
  Communications}, vol.~37, no.~4, pp. 854--868, 2019.

\bibitem{YZ2009}
Y.~Zhu, Q.~Zhang, Z.~Niu, and J.~Zhu, ``{Leveraging multi-AP diversity for
  transmission resilience in wireless networks: Architecture and performance
  analysis},'' \emph{IEEE Transactions on Wireless Communications}, vol.~8,
  no.~10, pp. 5030--5040, 2009.

\bibitem{giordani2019standalone}
M.~Giordani, M.~Polese, A.~Roy, D.~Castor, and M.~Zorzi, ``{Standalone and
  non-standalone beam management for 3GPP NR at mmWaves},'' \emph{IEEE
  Communications Magazine}, vol.~57, no.~4, pp. 123--129, Apr. 2019.

\bibitem{YN2015}
Y.~Niu, Y.~Li, D.~Jin, L.~Su, and A.~V. Vasilakos, ``{A survey of millimeter
  wave (mmWave) communications for 5G: Opportunities and challenges},''
  \emph{Journal on Wireless Networks}, vol.~21, no.~8, pp. 2657--2676, 2015.

\bibitem{GupAndHea2016}
A.~K. Gupta, J.~G. Andrews, and R.~W. Heath, ``On the feasibility of sharing
  spectrum licenses in mm{W}ave cellular systems,'' \emph{IEEE Transaction on
  Communication}, vol.~64, pp. 3981--3995, Sep. 2016.

\bibitem{AGAA2016}
A.~K. Gupta, A.~Alkhateeb, J.~G. Andrews, and R.~W. Heath, ``Gains of
  restricted secondary licensing in millimeter wave cellular systems,''
  \emph{IEEE Journal on Selected Areas in Communications}, vol.~34, no.~11, pp.
  2935--2950, Nov. 2016.

\bibitem{AA2018}
A.~Alkhateeb, S.~Alex, P.~Varkey, Y.~Li, Q.~Qu, and D.~Tujkovic, ``{Deep
  learning coordinated beamforming for highly-mobile millimeter wave
  systems},'' \emph{IEEE Access}, vol.~6, pp. 37\,328--37\,348, 2018.

\bibitem{TN2014}
T.~Nitsche, C.~Cordeiro, A.~B. Flores, E.~W. Knightly, E.~Perahia, and J.~C.
  Widmer, ``{IEEE 802.11 ad: Directional 60 GHz communication for
  multi-Gigabit-per-second Wi-Fi},'' \emph{IEEE Communications Magazine},
  vol.~52, no.~12, pp. 132--141, 2014.

\bibitem{YG2017}
Y.~Ghasempour, C.~R. da~Silva, C.~Cordeiro, and E.~W. Knightly, ``{IEEE 802.11
  ay: Next-generation 60 GHz communication for 100 Gb/s Wi-Fi},'' \emph{IEEE
  Communications Magazine}, vol.~55, no.~12, pp. 186--192, 2017.

\bibitem{YL2020}
Y.~Liu, Y.~Jian, R.~Sivakumar, and D.~M. Blough, ``{On the potential benefits
  of mobile access points in mmWave wireless LANs},'' in \emph{2020 IEEE
  International Symposium on Local and Metropolitan Area Networks
  (LANMAN}.\hskip 1em plus 0.5em minus 0.4em\relax IEEE, 2020, pp. 1--6.

\bibitem{KA2020}
K.~Aldubaikhy, W.~Wu, N.~Zhang, N.~Cheng, and X.~S. Shen, ``{Mmwave IEEE 802.11
  ay for 5G fixed wireless access},'' \emph{IEEE Wireless Communications},
  vol.~27, no.~2, pp. 88--95, 2020.

\bibitem{dhillon2015wireless}
H.~S. Dhillon and G.~Caire, ``Wireless backhaul networks: Capacity bound,
  scalability analysis and design guidelines,'' \emph{IEEE Transactions on
  Wireless Communications}, vol.~14, no.~11, pp. 6043--6056, Nov. 2015.

\bibitem{saha2018bandwidth}
C.~Saha, M.~Afshang, and H.~S. Dhillon, ``Bandwidth partitioning and downlink
  analysis in millimeter wave integrated access and backhaul for {5G},''
  \emph{IEEE Transactions on Wireless Communications}, vol.~17, no.~12, pp.
  8195--8210, Dec. 2018.

\bibitem{saha2019millimeter}
C.~Saha and H.~S. Dhillon, ``Millimeter wave integrated access and backhaul in
  {5G}: Performance analysis and design insights,'' \emph{IEEE Journal on
  Selected Areas in Communications}, vol.~37, no.~12, pp. 2669--2684, Dec.
  2019.

\bibitem{LR2019}
R.~Lombardi, ``{Wireless backhaul for IMT 2020 / 5G: Overview and
  introduction},'' in \emph{In Proceedings of the Workshop on Evolution of
  Fixed Service in Backhaul Support of IMT 2020/{5G}, Geneva, Switzerland}, 29
  April 2019.

\bibitem{MJ2016}
M.~Jaber, M.~A. Imran, R.~Tafazolli, and A.~Tukmanov, ``{5G backhaul challenges
  and emerging research directions: A survey},'' \emph{IEEE access}, vol.~4,
  pp. 1743--1766, 2016.

\bibitem{TSR2015}
T.~S. Rappaport, R.~W. Heath~Jr, R.~C. Daniels, and J.~N. Murdock,
  \emph{{Millimeter wave wireless communications}}.\hskip 1em plus 0.5em minus
  0.4em\relax Pearson Education, 2015.

\bibitem{SB2019}
S.~Barberis, D.~Disco, R.~Vallauri, T.~Tomura, and J.~Hirokawa, ``{Millimeter
  wave antenna for information shower: Design choices and performance},'' in
  \emph{2019 European Conference on Networks and Communications (EuCNC)}.\hskip
  1em plus 0.5em minus 0.4em\relax IEEE, 2019, pp. 128--132.

\bibitem{SJ2019}
S.~Jaswal, D.~Yadav, D.~P. Bhatt, and M.~Tiwari, ``{MmWave technology: An
  impetus for smart city initiatives},'' 2019.

\bibitem{dhillon2020poisson}
H.~S. Dhillon and V.~V. Chetlur, \emph{Poisson Line Cox Process: Foundations
  and Applications to Vehicular Networks}.\hskip 1em plus 0.5em minus
  0.4em\relax Morgan \& Claypool, Jun. 2020.

\bibitem{PK2015}
P.~Kumari, N.~Gonzalez-Prelcic, and R.~W. Heath, ``{Investigating the IEEE
  802.11 ad standard for millimeter wave automotive radar},'' in \emph{2015
  IEEE 82nd Vehicular Technology Conference (VTC2015-Fall)}.\hskip 1em plus
  0.5em minus 0.4em\relax IEEE, 2015, pp. 1--5.

\bibitem{JH2012}
J.~Hasch, E.~Topak, R.~Schnabel, T.~Zwick, R.~Weigel, and C.~Waldschmidt,
  ``{Millimeter-wave technology for automotive radar sensors in the 77 GHz
  frequency band},'' \emph{IEEE Transactions on Microwave Theory and
  Techniques}, vol.~60, no.~3, pp. 845--860, 2012.

\bibitem{YH2016}
Y.~Han, E.~Ekici, H.~Kremo, and O.~Altintas, ``{Automotive radar and
  communications sharing of the 79-GHz band},'' in \emph{Proceedings of the
  First ACM International Workshop on Smart, Autonomous, and Connected
  Vehicular Systems and Services}, 2016, pp. 6--13.

\bibitem{VP2018}
V.~Petrov, J.~Kokkoniemi, D.~Moltchanov, J.~Lehtom{\"a}ki, M.~Juntti, and
  Y.~Koucheryavy, ``{The impact of interference from the side lanes on
  mmWave/THz band V2V communication systems with directional antennas},''
  \emph{IEEE Transactions on Vehicular Technology}, vol.~67, no.~6, pp.
  5028--5041, 2018.

\bibitem{VP2019}
V.~Petrov, G.~Fodor, J.~Kokkoniemi, D.~Moltchanov, J.~Lehtomaki, S.~Andreev,
  Y.~Koucheryavy, M.~Juntti, and M.~Valkama, ``{On unified vehicular
  communications and radar sensing in millimeter-wave and low terahertz
  bands},'' \emph{IEEE Wireless Communications}, vol.~26, no.~3, pp. 146--153,
  2019.

\bibitem{Akyildiz2014}
I.~F. Akyildiz, J.~M. Jornet, and C.~Han, ``{Terahertz band: Next frontier for
  wireless communications},'' \emph{Journal of Physics Communication}, vol.~12,
  pp. 16--32, Sep. 2014.

\bibitem{Singh2019}
R.~Singh and D.~Sicker, ``{Parameter modeling for small-scale mobility in
  indoor THz communication},'' in \emph{Proceedings of GLOBECOM}.\hskip 1em
  plus 0.5em minus 0.4em\relax IEEE, Dec. 2019, pp. 1--6.

\bibitem{Kurner2014}
T.~K{\"{u}}rner and S.~Priebe, ``{Towards THz communications - Status in
  research, standardization and regulation},'' \emph{Journal on Infrared,
  Millimeter and Terahertz Waves}, vol.~35, no.~1, pp. 53--62, Jan. 2014.

\bibitem{Saeed2020}
A.~Saeed, O.~Gurbuz, and M.~A. Akkas, ``{Terahertz communications at various
  atmospheric altitudes},'' \emph{Journal on Physics Communication}, vol.~41,
  pp. 101--113, Aug. 2020.

\bibitem{Rasheed2020}
I.~Rasheed and F.~Hu, ``{Intelligent super-fast vehicle-to-everything 5G
  communications with predictive switching between mmWave and THz links},''
  \emph{Vehicular Communication}, p. 100303, Sep. 2020.

\bibitem{Akyildiz2015}
I.~Akyildiz, M.~Pierobon, S.~Balasubramaniam, and Y.~Koucheryavy, ``{The
  internet of bio-nano things},'' \emph{IEEE Communication Magzine}, vol.~53,
  no.~3, pp. 32--40, Mar. 2015.

\bibitem{Abadal2013}
S.~Abadal, E.~Alarc{\'{o}}n, A.~Cabellos-Aparicio, M.~Lemme, and M.~Nemirovsky,
  ``{Graphene-enabled wireless communication for massive multicore
  architectures},'' \emph{IEEE Communication Magzine}, vol.~51, no.~11, pp.
  137--143, Nov. 2013.

\bibitem{NAF2019}
N.~Al-Falahy and O.~Y. Alani, ``{Millimetre wave frequency band as a candidate
  spectrum for 5G network architecture: A survey},'' \emph{Journal of Physics
  Communication}, vol.~32, pp. 120--144, 2019.

\bibitem{PZ2018}
P.~Zhou, K.~Cheng, X.~Han, X.~Fang, Y.~Fang, R.~He, Y.~Long, and Y.~Liu,
  ``{IEEE 802.11 ay-based mmWave WLANs: Design challenges and solutions},''
  \emph{IEEE Communications Surveys and Tutorials}, vol.~20, no.~3, pp.
  1654--1681, 2018.

\bibitem{JP2020}
J.~Peisa, P.~Persson, S.~Parkvall, E.~Dahlman, A.~Grovlen, C.~Hoymann, and
  D.~Gerstenberger, ``{5G evolution: 3GPP releases 16 \& 17 overview},''
  \emph{Ericsson Technology Revisions}, vol.~9, pp. 1--5, 2020.

\bibitem{molstd}
``{IEEE P1906.1/Draft 1.0},'' \emph{Recommended practice for nanoscale and
  molecular communication framework}, 2014.

\bibitem{Elayan2020}
H.~Elayan, O.~Amin, B.~Shihada, R.~M. Shubair, and M.-s. Alouini, ``{Terahertz
  band: The last piece of RF spectrum puzzle for communication systems},''
  \emph{IEEE Open Journal of Communication Society}, vol.~1, no. Nov., pp.
  1--32, 2020.

\end{thebibliography}

\end{document}